# Deconstructing Sunlight – A Community Enterprise

**J. W. Harvey**
National Solar Observatory, Boulder, CO USA
Lunar and Planetary Laboratory, University of Arizona, Tucson, AZ USA
e-mail: jharvey@nso.edu

**Abstract** Good fortune allowed me to have a career in solar research. With the aid of many generous and gifted colleagues, I've tried to learn more about the Sun's magnetic and velocity fields through observation and instrumentation development. These interests captured me early and remain strong. In this memoir I describe my path through 60 years of solar research that was sometimes random but did not deviate much from my core interests. The chromospheric magnetic field and helioseismology have been especially intriguing and frequently rewarding topics.

## 1. Introduction

"He was a strange child. I hardly knew what to do with him." my mother said to my now wife, Terrie, when they first met. Even so, through good fortune, generous people, and being at the right place at the right time many times, I've had an unusually rewarding career. This memoir conveys the experiences of one lucky person fascinated by the Sun during the second half of the 20th century. The emphasis is on the earlier years of my career. Conditions then were much different than now. With changes in society, many of my favorable experiences cannot be repeated in contemporary times. However, there may be some enduring principles exposed here of some interest to younger scientists. I've tried to be accurate but old memories are a fallible record of reality. I name a lot of people here because without them, and many unnamed others, I would not have had my privileged career.

My scientific career story stands on a framework of people and environments that provide structure to this memoir: Parental support, early education, amateur astronomy, university education, Lockheed Solar Observatory, Mount Wilson Observatory, High Altitude Observatory, life with Karen, Kitt Peak National Observatory (later National Solar Observatory), and life with Terrie. My research interests formed early and most have persisted throughout my career: synoptic observations, flare activity, magnetic field measurements (especially in the chromosphere), coronal holes, and velocity measurements (especially helioseismology).

## 2. Origins

I was born in Hollywood, California, 15 months before the United States entered World War II. My parents were from small farming towns in northeast Nebraska in the central part of the United States, and they experienced first-hand the harrowing Great Depression of the 1930s. My father started work as a salesman and manager of his family's farming-town hardware store. He had interests and talents in photography, electronics, graphic arts, and music. A childhood accident significantly impaired his vision. My mother was a farmer's daughter, intelligent and skilled in home crafts. She was headed to a career as a law secretary when she met and married my father in 1937. My parents did not have the luxury of formal education beyond solid high school basics, but they shared lifelong loves of learning, great curiosity, and a sense of adventure. Like many, they sought a better life in more prosperous Southern California. Within their limited financial resources, they were incredibly supportive of my younger sister and me as we grew up. An early photo of me (Figure 1) suggests that listening to tedious lectures might be in my future. Post-war economic growth in the United States was far more equitable than today's economy, and our family became part of a growing lower middle class. The third quarter of the 20$^{th}$ century was an unusually favorable economic period in the United States. In 1950 we moved to a small house in North Hollywood in the booming San Fernando Valley of Southern California. That move proved to be very lucky for me and my career yet to come.

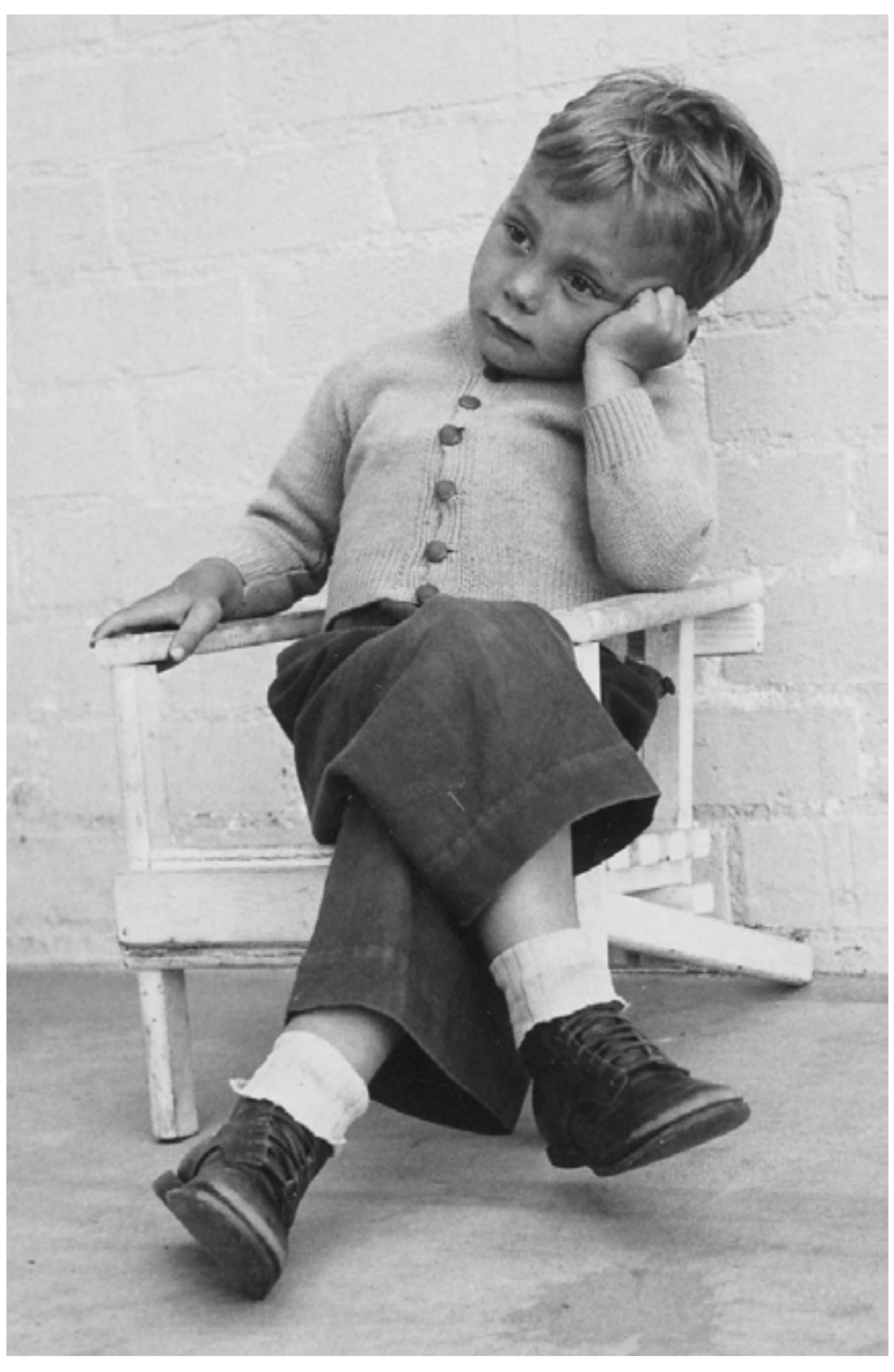

**Figure 1** Contemplating a career in science in 1944?

My interest in astronomy was sparked by my father, who had interests in technology and with whom I saw a still-remembered meteor shower on October 9, 1946. Family trips to Griffith Observatory and Planetarium in Hollywood, to Mt. Wilson and Mt. Palomar Observatories, and excellent public school teachers strengthened my fascination with science in general and astronomy in particular. I read and reread popular astronomy books and magazines of the day and wanted a telescope to see celestial wonders myself. My father's job as a professional photographer[1] brought him into contact with an early television shopping program and he obtained a small 10X draw-tube achromatic telescope, which I still have. With this I started to explore telescope optics and celestial objects. Less powerful than Galileo's telescope, I was still able to see much of what he had discovered centuries earlier. One of our family trips was to the Hale 200-inch telescope on Mt. Palomar with no inkling that Hale and his legacies would play a big role in my future (Figure 2).

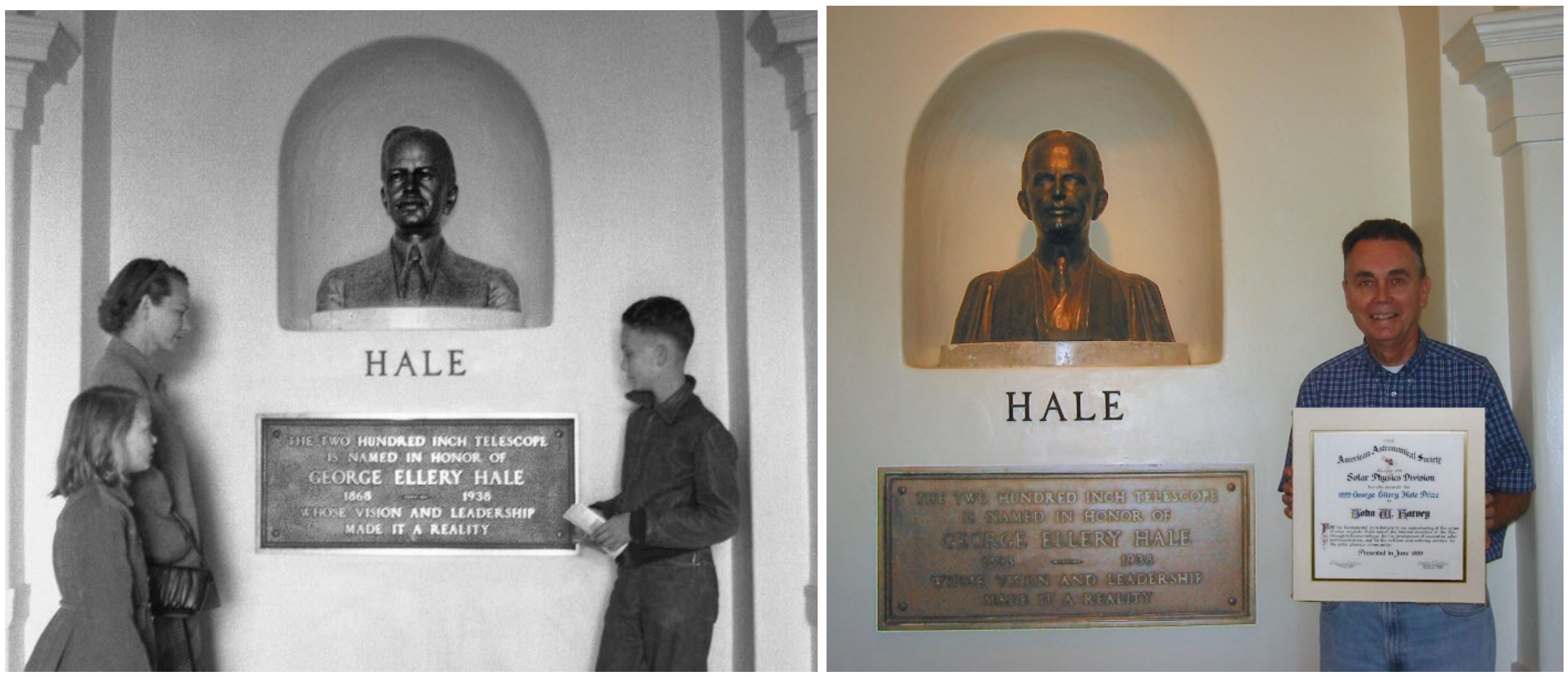

**Figure 2** (Left) My sister Carol, mother Laureene, and me in the entrance to the visitor's gallery of the Hale Mt. Palomar 200-inch telescope, ca. 1951. (Right) Half a century later I received the Hale Prize and Medal of the American Astronomical Society and revisited the same location.

## 3. Early Education and Amateur Astronomy

When I was about age 10 my parents gave me a collection of surplus chipped lenses purchased as a kit from Edmund Salvage Company (still available from Edmund Optics!) I experimented with the lenses a lot, not knowing at the time that my future career was being set. After starting junior high school, I met neighborhood boys who had similar interests in science and astronomy. We formed a club and built our own telescopes with each other's help, and became avid amateur astronomers. I also attended meetings of the Los Angeles Astronomical Society where I met and photographed famed Edwin Hubble, who lectured about the red shift of galaxies shortly before his death. In those days the economical way to get a good telescope was to build it yourself. We learned many of the necessary skills in junior high school industrial arts classes in metal work and electronics among other subjects. I ground and polished five telescope mirrors in all and pretty much ruined my mother's kitchen in the process.

[1] http://www.csun.edu/bradley-center/bill-harvey

Our club heard about a solar observer at Mt. Wilson Observatory who enjoyed showing young people around the observatory, and so we arranged a trip. The observer was Joseph Hickox[2] and he became one of my many mentors. Figure 3 shows me and friends Kent and Bill with Joe pointing out sunspots on the solar image at the 150-foot solar tower on Mt. Wilson. This visit and Joe's warm welcome focused me on the road to solar astronomy. Subsequent visits only increased my intention to try to become a solar astronomer. In retrospect, I was unbelievably naïve about this, having only a vague idea how to become an astronomer or how few jobs were actually available in astronomy and especially solar astronomy. But I continued to observe the Sun at home and also many nighttime wonders in dark sky locations with my astronomy club friends. I built a heliostat and spectrograph and was thrilled to see prominences above the solar limb using the Hα line and a wide entrance slit. My close friends Barry Nolan[3], Kent DeGroff[4], Denis Dutton[5], and Jerg Jergenson[6] had related interests and we educated each other in many ways to do and enjoy observational astronomy.

---

[2] http://hdl.huntington.org/cdm/search/searchterm/Hickox,%20Joseph./mode/exact
[3] http://www.imdb.com/name/nm0634219/

[4] https://www.flickr.com/photos/whiskey_creek_observatory/
[5] https://en.wikipedia.org/wiki/Denis_Dutton
[6] http://patents.justia.com/inventor/jerg-b-jergenson

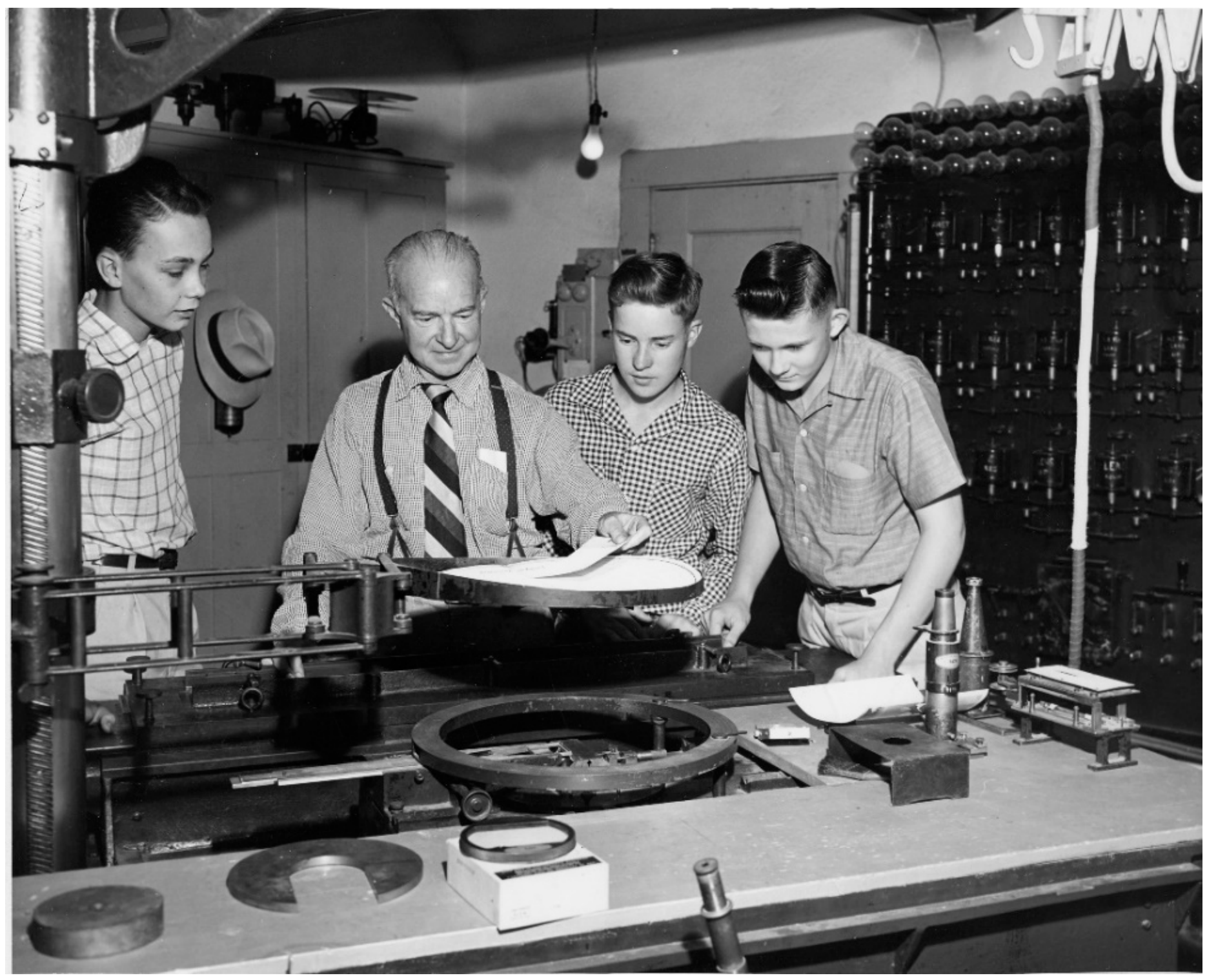

**Figure 3** (Left to right) Me, veteran solar observer Joseph O. Hickox, friends Kent DeGroff and F. W. Milburn in the observing room of the 150-foot solar tower at Mt. Wilson Observatory in 1954.

In addition to the valuable practical shop experiences at North Hollywood Junior High School (now Walter Reed Middle School), I took all the classes that seemed appropriate for a career in science. My vague backup plan was aerospace engineering. When I graduated from junior high school in 1955 I was asked to give a speech at the ceremony. Being shy, I didn’t want to do it, and so I drafted a very short talk in the hope that someone else would be picked. No such luck, but I learned a valuable lesson that has served me well. Namely, audiences appreciate short talks more than long ones. Much later, Art Hundhausen further educated me by saying, “Aim low. If the listeners already know the subject, they will feel smart. If they don’t, they will appreciate simple explanations.”

At North Hollywood High School, I was again lucky to have excellent math, chemistry, and physics teachers who were willing to spend after-school time with eager students. Electronics became a hobby through the construction of various inexpensive kits, mainly from the now defunct Heathkit Company. One of the kits was a small oscilloscope. I also built a variable frequency oscillator from scratch that was valuable for controlling tracking of my telescopes. I experimented with the newly invented transistor (the stupendous exponential growth of solid state electronics astounds me with $8 \times 10^{12}$ transistors produced per *second* in 2014 (Hutcheson, 2015)). I continued to take classes that prepared me for college. Our small group of amateur astronomers grew larger (Harvey, 1957), and we went on “star parties” to dark skies north of Los

Angeles to observe and photograph comets and nebulae. Viewing Comet Arend-Roland was particularly exciting in early 1957 as was the first *Sputnik* satellite later that year. Learning chemical photographic techniques from my father was a lot of fun in those pre-digital days and served me well into the 1970s.

My father sometimes invited me along on his commercial and publicity photography jobs. On one occasion he said that I might want to go on a job with him because of my growing interest in classical music but he didn't say where we were going. I was thrilled when it turned out to be famed composer Igor Stravinsky's home. During these mid-teen-age years I took various jobs to earn money, including a brief stint working at Disneyland. I built a commissioned telescope and mounting for a physician for a sale price of $100. My friend Barry Nolan became very interested in motion picture animation and found a way to sneak into the Burbank Walt Disney Studios where some of the animators had befriended him. I accompanied one of his illegal entries and needlessly feared being arrested the whole time. Since my father was involved in publicity, I was exposed to that curious industry and (reluctantly) participated in it. For example, Barry had used one of my telescopes to make a movie of various regions of the moon and we both appeared on a local television program to show the movie and talk about astronomy.

## 4. UCLA 1958-1963

Once a year, the California Institute of Technology (Caltech) hosted an open house for selected high school science students from all over the Los Angeles area. My chemistry teacher nominated me, and I heard an inspiring lecture by Linus Pauling and recall seeing a lab setup of the then relatively new nuclear magnetic resonance phenomenon and other fascinating research in progress. Though Caltech was well-connected with the solar program at Mt. Wilson, I did not apply to go to college there because of well-founded fears of personal financial and academic inadequacies. Unlike today, as a resident of California, the cost to attend a state-supported university was then affordable even for the lower middle class, consisting of an annual fee of $300 and no tuition (http://www.dailycal.org/2014/12/22/history-uc-tuition-since-1868/). So I applied to and was accepted at the University of California at Los Angeles (UCLA). As an undergraduate in the Astronomy Department I quickly learned that I was surrounded by much smarter people. While I would have had a fatal time at Caltech, it was hard enough at UCLA and I proved not to be an academic standout. During my first year, one of my classes was a mathematics course taught by Edward O. Thorp. Oddly, Thorp used numerous examples involving casino gambling, which seemed strange. It turned out that he was then developing his successful technique for winning the casino game blackjack (Thorp, 1962). He continued his mathematical success by studying the stock market and in 2012 had a reported net worth of nearly a billion dollars. I should have paid more attention in his class!

As a so-called land grant university it was required that UCLA male students participate in a military program called reserve officer training corps (ROTC). I did this for my first year after which the ROTC requirement was eliminated. During my college years the Vietnam War raged and there was a possibility of being drafted. Early on it was relatively easy to obtain a student deferment, but this became more difficult as the war expanded. For some time periods I was not deferred but was never drafted. One reason might have been that an unknown war protestor destroyed the draft office that held my records. By the time my academic career was finished, I was too old to be militarily useful.

Most of the astronomy courses that I took as an undergraduate proved to have lasting value in my career. These included at least one course each by professors F. C. Leonard, L. H. Aller, G. O. Abell, and D. M. Popper who constituted the UCLA astronomy faculty in those days. The small department was strong in many areas of astronomy but was weak in solar physics and tainted by a then-common bias against women. To indicate the success of the department, several of my fellow students later became well-known astronomers including David DeVorkin[7], Don Goldsmith[8], Don Hayes[9], Steve Little[10], Joe Miller[11], Gerrie Peters[12], Stephen Price[13], Sumner Starrfield[14], Virginia Trimble[15], and so on.

The UCLA astronomy department offered part-time employment to its undergraduate students as "readers" for 4 hours per week at $1.90 per hour. I welcomed this financial opportunity. In practice the work involved helping professors on their research projects and grading introductory astronomy homework and exams. One typical example was recently published by my former office mate at UCLA, David DeVorkin (2015). The department had a number of Fridan mechanical calculators that we used to crank out calculations for our courses. But, becoming fascinated with digital computing, I took an introductory computer course. UCLA had inherited the tenth vacuum-tube computer ever built, called SWAC, which, when built in 1950, was for one year the fastest computer in the world (https://en.wikipedia.org/wiki/SWAC_(computer)). By the 1960s it was mainly an instructional tool. Novel was the fast 256-word memory of 37-bit words stored within cathode-ray tubes. One could see every single bit in the memory rendered as ones or dots by looking at the glowing screens of the 256 tubes. I learned binary logic, assembly language, and Fortran II in that course. Later courses and experience improved my programming, as did access to UCLA's IBM 7094 scientific computer, quite advanced for its time.

My undergraduate academic career was undistinguished, in part because of too much involvement with observatories as described below. It seemed that my formal education might end after my graduation in early 1963. However, encouraged by my mentors at Lockheed (see Section 5) and Prof. Popper, I decided to study for a Master's degree with additional courses and by doing an observational project at Mt. Wilson and Lockheed on solar chromospheric magnetism (a subject I'm still working on 57 years later). I worked half-time for the astronomy department as a teaching assistant for George Abell – a brilliant researcher and inspiring professor. My project work was done during 1963 and early 1964. Prof. Popper was not an expert in solar work, and so he gave my 47-page project report for review to Charles Hyder, who had recently joined the department as an assistant professor. Charlie wrote a positive review that was almost as long as my report. I graduated in June 1964, feeling more confident in my career plans.

[7] https://airandspace.si.edu/staff/david-devorkin
[8] http://www.asroc.org.tw/asroc2013/en/DonaldGoldsmith.php
[9] https://www.aip.org/history-programs/niels-bohr-library/oral-histories/4660
[10] http://www.iop.org/careers/workinglife/articles/page_39064.html
[11] https://news.ucsc.edu/2005/10/753.html
[12] https://dornsife.usc.edu/cf/phys/faculty_display.cfm?person_ID=1008400
[13] https://users.physics.unc.edu/~gcsloan/library/2012/price/
[14] http://starrfield.asu.edu/
[15] https://en.wikipedia.org/wiki/Virginia_Louise_Trimble

## 5. Lockheed Solar Observatory 1959-1968

I needed to earn some money during the 1959 summer break after my first undergraduate college year. I'd heard about a new solar observatory that had been set up by Lockheed Aircraft Corporation headquartered in nearby Burbank. In early post-*Sputnik* days, the United States government greatly increased funding for science, especially for anything related to space. One result was an abundance of resources to help set up new research facilities, and major aero-space corporations seized this opportunity to expand beyond their defense and commercial activities. At that time Lockheed's Chief Scientist and director of a research staff of more than 400 was solar astronomer Lewis Larmore, whose 1952 UCLA PhD thesis dealt with solar prominences (Larmore, 1953). This was based on his 1944-1946 duty as a naval officer stationed at Harvard College Observatory's Fremont Pass Station at Climax, Colorado, working with Walt Roberts. I'm not certain if Lew initiated the Lockheed Solar Observatory project or if it had been proposed to him by Gail Moreton[16]. I suspect the former. Regardless, I contacted Gail who was in charge of the observatory project and asked if there might be a summer job opportunity. He invited me for a tour and interview. Probably because of my amateur observing and telescope building experiences and familiarity with photographic techniques, he offered me a job. I was at the right place at the right time.

The Lockheed observatory became operational in 1958 during the latter part of the International Geophysical Year (IGY). It was located at Lockheed's flight test radio station on Briar Summit in the Hollywood Hills southwest of Burbank. The original Climax coronagraph that Larmore had used for his thesis work had been loaned to Lockheed by the High Altitude Observatory. It was equipped with a 3-inch aperture f/33 lens, a Bernard Halle 0.5 Å H-alpha filter, and a 35-mm Acme/Photo-Sonics single-frame instrumentation 35 mm film camera. The observatory was part of the IGY world-wide H-alpha flare patrol network. The history of the observatory is described in several publications (Moreton, 1960, 1962; Smith, 1963; Anderson, 1967; Acton, 1969a, b, 1970; Nolan, Smith, and Ramsey 1970; Martin 2015). My initial job was to help operate the flare patrol, develop film taken with the instrument, review the film, and record descriptions of various solar activity. I started work on 13 July 1959. That date was fortunate since one of the largest flares of cycle 19 occurred three days later while I was the novice operator at the observing site. A then unique aspect of the Lockheed flare patrol was a fast cadence of one frame per 10 s instead of the prevailing one frame every few minutes. This made it hard to sneak looks at the image without blocking the film camera, but I still have a vivid recollection of the dazzling brilliance of that very large flare.

[16] http://answers.google.com/answers/threadview/id/283931.html

The staff was small when I joined, consisting of Gail Moreton, Neil Christie, George Carroll, and me. Later that summer Gerry Anderson (a graduate study scientist invited by Lew Larmore from the High Altitude Observatory) briefly joined the group. Neil Christie taught me how to operate the flare patrol and the basics of analyzing the film records for interesting events. He did the preparation of data reports using IBM equipment – my first introduction to the world of digital data base handling. George Carroll was an aircraft engineer in Lockheed's Advanced Development Projects and a brilliant designer and builder of astronomical instruments and telescopes (Briggs, 2020). It is indeed fortunate that George contributed his skill to the instrumental capabilities of the observatory. A major staff addition in September 1959 was veteran Sacramento Peak Observatory observer Harry Ramsey[17].

When I started working at Lockheed, we shared desks on the top floor of a large engineering building. Here aircraft flight test data recorded on film was developed and analyzed by a large staff located in a big room with many IBM key punch machines and Richardson film readers (Clark and Richardson, 1951). The room was noisy but was well equipped to study and report the flare patrol films. A departmental address system provided frequent, loud announcements of no interest to observatory work. To cope with this, Gail stuffed tissues in the ceiling loudspeaker above our desks. (Gail wanted to be addressed as "GE" but we always called him Gail.) Later that summer we moved into adjacent, dedicated, and quieter rooms in the same building. One reason might have been security, since the room we originally occupied handled some data from classified projects. Later I learned that we were adjacent to Lockheed's Advanced Development Projects building, more commonly known as the Skunk Works. We were aware of the U-2 and F-104 aircraft projects but (likely except for George) had absolutely no knowledge of the A-12/SR-71 development going on next door. I was incredibly lucky to be mentored by Harry, Gail, and George and for early exposure to life in a big corporation.

The observatory was thriving under Gail Moreton's entrepreneurial leadership and support from Lockheed's management. Though lacking the usual academic credentials, he energetically promoted the work of the observatory, locally, then nationally, and finally internationally at meetings and by inviting well-known solar astronomers to visit and collaborate. As a result, I met many famous solar astronomers very early in my career. One example is Figure 4.

[17] See Ramsey, Joanne 1997, In The Beginning, Morris Publishing, Kearney Nebraska.

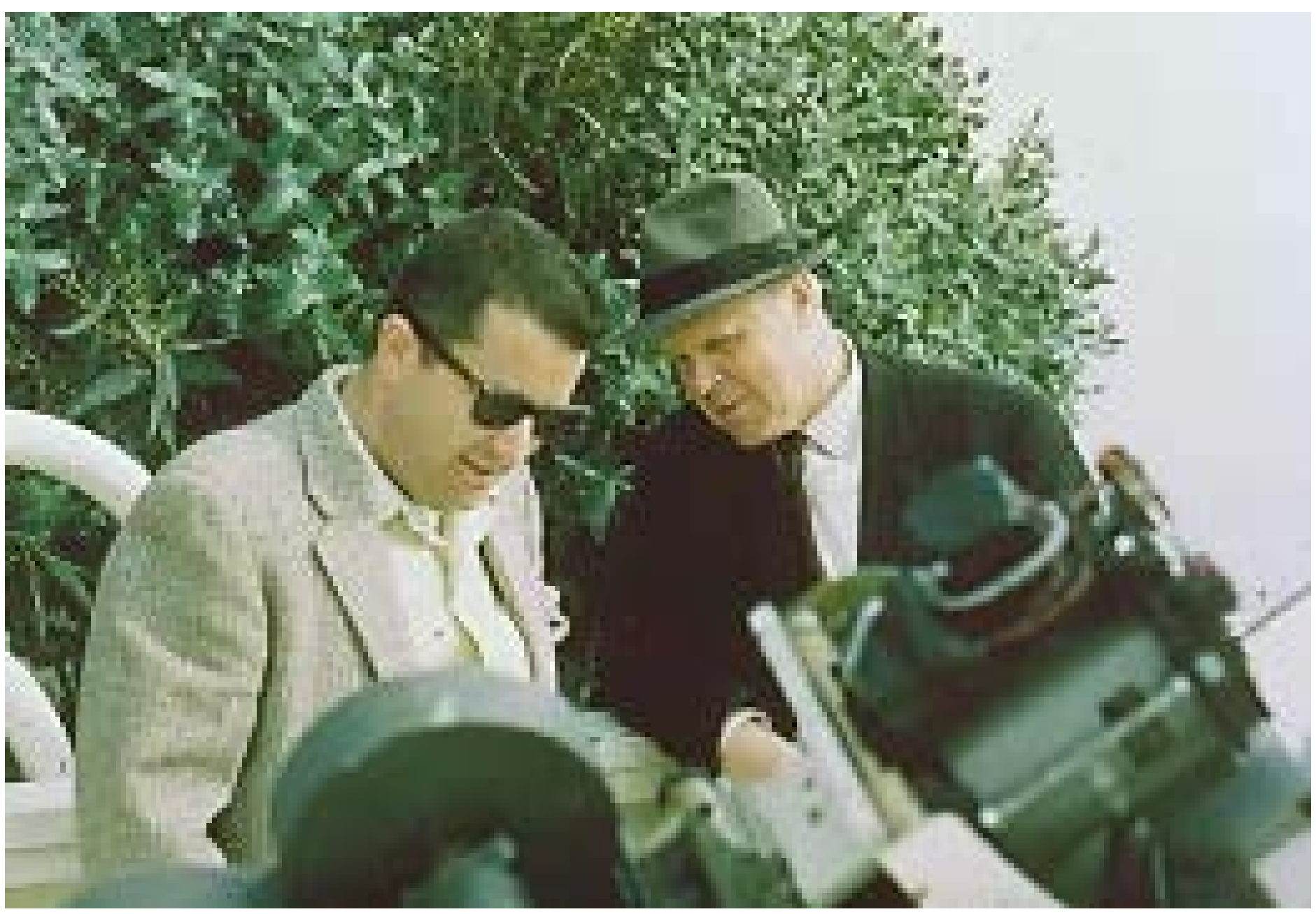

**Figure 4** Gail E. Moreton (left) showing visiting professor Yngve Öhman (right) the Lockheed Solar Observatory flare patrol on Briar Summit. In the lower right foreground is a Lyot-Öhman filter of the type Öhman was the first person to develop. A motor on top of the filter tuned it alternately between blue and red Hα line wings.

The film developing machine at Lockheed used a powerful developer that was not well matched to the flare patrol film. And the machine frequently broke down. Being so close to Hollywood's film processing labs, Harry Ramsey suggested that we use one of these commercial labs. We had already used one of the labs to make 16-mm movie prints of interesting events, and adding the routine film development to our account was easy. It was delightful for me to no longer have to deal with a cranky developing machine. Another inspired innovation of Harry's was to switch the wavelength of the Hα filter to either the red or blue wing of the line, something he had long wanted to do as an observer at Sacramento Peak.

After returning to UCLA for my second academic year, both for the money and because I was interested in the work, I continued working at Lockheed part time by opening and closing the flare patrol, transporting exposed film to the processing lab, and assisting with studies of flare-related disturbances that had been revealed by the fast cadence of the observations. This was my first (and very lucky) participation in real solar research, although my role was tiny. A notable observation on the flare patrol films was that a few filaments would sometimes suddenly and briefly disappear or greatly weaken shortly after a nearby flare (Moreton, 1961). When viewed as movies, some of the events looked as if an unseen force was traveling along the chromosphere between the flare and the filament. It was not long until Harry's foresight in running the patrol using the wing of the Hα line soon revealed what the line-center observations did not.

When I returned to full time work at Lockheed during the summer of 1960, the staff roster was unchanged and interest in the mysterious effect of flares on distant filaments was a strong research focus together with so-called impulsive flares. Sometimes when viewing the patrol films with the Richardson film viewer, we had the impression of a faint disturbance moving outward at high speed away from a flare site. But the viewing machine would not project the frames very rapidly, and the projected image contrast was not high. Multiple attempts to see the faint motions at maximum machine speed also tended to damage the film. Thanks to Harry's off-band insight a particularly intriguing event was recorded in June. I recall going to the film processing lab to pick up a 16-mm movie print from the 35-mm original flare patrol negative and taking it to our office. Gail, Harry, and I put the movie in a projector, turned down the lights, and projected the film. It was a "Wow!" moment when we saw a clear wave-like disturbance moving away from the flare. What a thrill for a young guy like me. Gail grabbed anyone walking by to show them this discovery, later dubbed a Moreton wave (Moreton and Ramsey, 1960).

To visualize and measure the still-faint moving disturbance, Harry introduced me to the technique of photographic image subtraction. Today, we take image subtraction for granted as a powerful method for revealing motions and subtle changes of faint solar features. But in 1960 it was non-trivial to do this using film. The key was to make a positive contact copy of a negative image using a certain film type and development so that when sandwiched together the pair would show only a high-density, uniform background. When sandwiched with a negative taken at a different time, any differences would be revealed. With this tedious technique a difference movie could be made that clearly showed the traveling disturbance and its position could be measured on single frames (Moreton, 1961).

After the exciting 1960 summer, I continued toward my bachelor's degree while working part time at Lockheed. During the following summer of 1961, I again worked full time at Lockheed. The staff was the same plus the addition of Sam Miller[18], a Weather Bureau statistician, as another part-time observer. We continued to work on Moreton waves and other flare phenomena. For some now-forgotten reason Gail decided to send me to the 1961 IAU General Assembly in Berkeley to hear the latest in solar research. That was an exciting meeting with early reports about the discovery of solar oscillations and Babcock's presentation of his dynamo model, but I felt really out of place as an undergraduate among world leaders in astronomy. In September 1961, I returned to part-time work at Lockheed, and Sara Smith (now Sara Martin) joined the Lockheed group that month to strengthen its research capabilities (Martin, 2015).

I worked full time at Lockheed during the summer of 1962, part time in 1963, and full time in 1964. I continued my observations of the chromospheric magnetic field in part by using a newly-available, red-sensitive photomultiplier sensor purchased by Lockheed. Major changes took place as the solar group moved to a nicely-equipped, new research campus at Rye Canyon, near Saugus, California. The group grew with the addition of Don Carson, Larry Stoddard, and later my boyhood friend Barry Nolan as a part timer. Gail Moreton's job was terminated as a result of his failure to get prior approval before traveling, and the consequent disagreement with Lockheed management. The resulting management vacuum was filled by Harry Ramsey as reported by Martin (2015). Harry and Sara sustained the observatory through a series of successful research proposals to various research funding agencies. On one occasion the president of Lockheed brought Edward Teller to visit the lab facilities. Ever eager to show work in progress and little aware of reputations, I remember Harry saying, "Did you say your name was Ed? Ed, let me show you this." It turns out that Teller did not like to be called Ed, but the visit was nevertheless successful and as a result I met another famous person. I continued occasional part-time work at Lockheed as a consultant from 1965 until 1968 when the observatory moved away from Southern California.

[18] http://cloud-maven.com/trace-king-how-to-be-one/

## 6. Mt. Wilson Observatory 1961-1969

It is my uncertain recollection that I first became professionally involved with the solar program at the Mount Wilson and Palomar Observatories (MWO) in 1961 when at Lockheed we needed to examine the transmission profile of a Hα filter using a good spectrograph. Because of my boyhood visits to Mt. Wilson hosted by Joe Hickox, I knew that there was a very good spectrograph at the 150-foot tower telescope. Robert Howard was then heading the Mt. Wilson solar program and he allowed us to set up the filter and use the spectrograph. In 1962, I approached Bob Howard about doing joint observations with the magnetograph at Mt. Wilson and with a new 18 cm aperture Hα telescope just added to the Lockheed solar equipment. This led to my first co-authored paper in the *Astrophysical Journal* (Howard and Harvey, 1964).

During the summer of 1963, after I completed a bachelor's degree from UCLA, Lockheed sent a large total solar eclipse expedition to Alaska and I was included thanks to Lew Larmore. I was working part time at Mt. Wilson and Lockheed that summer and proposed a joint coronal plume eclipse project to Bob Howard. Lacking a suitable telescope, in a frantic two-week period I ground and polished a 15 cm f/5 mirror, and mounted it as a Newtonian reflecting telescope with a 35 mm camera at the focus. The pictures turned out nicely. The American Astronomical Society (AAS) scheduled its annual summer meeting in conjunction with the eclipse, so I became a junior member of the AAS and presented my first paper about chromospheric magnetic field observations (Harvey, 1963). I was honored that my AAS membership application was endorsed by MWO staff members Bob Howard and Alfred Joy. At that meeting the AAS met in one lecture room, and all the attendees could hear all of the talks. There were no posters or parallel sessions. At one meal event I happened to be seated at a table with a group who introduced themselves. I was astonished and thrilled when one older gentleman introduced himself to the table: Ejnar Hertzsprung, of H-R Diagram fame.

The 1963 eclipse was beautiful even if slightly affected by thin cirrus clouds. The goal of my project with Bob Howard was to test a presumption that coronal polar plumes are associated with magnetic field concentrations on the solar surface. As a surrogate for magnetic fields we used Ca

K line emission features in an image taken at Mt. Wilson at the same time as the eclipse in Alaska. The initial results were rather weak but did not contradict an association (Harvey, 1965).

In 1962, Bob started a program using the 150-foot tower to study the then recently-discovered 5-minute solar oscillations (Howard, 1967). My part-time summer 1963 job at MWO was to help make the observations for Bob's program. Observing with this facility was my tender introduction to what later became a big part of my career – helioseismology. I learned how to make observations with the Babcock magnetograph, which enabled me to make digital maps of the chromospheric magnetic field for my UCLA Master's degree project. I also assisted with a joint Mt. Wilson – Lockheed project to study the evolution of active region magnetic fields (Bumba and Howard, 1965a), and of background fields (Bumba and Howard, 1965b) by drawing isogauss contour maps from the Mt. Wilson magnetogram archive and constructing synoptic solar rotation maps (Howard, Bumba, and Smith, 1967). In the summer of 1965 I returned for my last employment at MWO but I continued working with Bob Howard informally for the duration of my doctoral work. Digitized observations with the MWO magnetograph allowed us to analyze full-disk Doppler-shift maps, which, due to my doctoral work, were delayed for a few years (Howard and Harvey, 1970).

## 7. High Altitude Observatory and University of Colorado 1964-1969

Following my graduation from UCLA with a Master's degree in June 1964, I applied for admission to the doctoral program in the Astro-Geophysics (AG) Department at the University of Colorado (CU). At that time, CU and Caltech were the leading institutions for observational solar research in the US. Luckily, in spite of my unspectacular academic record, I was admitted and moved to Boulder in September 1964. I suspect that Hal Zirin, who was familiar with me and my work at Mt. Wilson, was a key to my admission. Other behind-the-scenes supporters probably helped, too. In those days there was a close association between the High Altitude Observatory (HAO), CU, and Sacramento Peak Observatory (SPO). The US Air Force operated SPO and provided funding to needy doctoral students at CU for their work at HAO, which I was luckily offered and gratefully accepted.

During my first years at CU/HAO, I was required to take physics courses such as Quantum Mechanics and Magnetohydrodynamics as well as various solar-related courses in the AG department. Two foreign language reading proficiency tests were also mandated. I had taken French for several years, and enough German to pass the tests. Like my mother, my advisors hardly knew what to do with me. I passed through the hands of advisors John Firor, Julius London, Jack Eddy, Gordon Newkirk, and finally Einar Tandberg-Hanssen. All of these folks taught fine courses, but I especially liked Jack Eddy's course on solar physics. He was a superb teacher. My assigned HAO work for pay mainly involved background studies for an Advanced Orbiting Solar Observatory proposal, design and construction of a large clean air tunnel to test coronagraphs, and some optics assembly for the Newkirk White Light Coronal Camera first used in 1966 (http://mlso.hao.ucar.edu/hao-eclipses.php). This work was my first real involvement with the growing world of space solar physics. I shared a CU campus office in Sommers-Bausch Observatory with other AG graduate students. These included Ernie Hildner, Robert Davies-Jones, Dan Tarpley, Christopher Kaiser, and Abdul Majid, who all went on to distinguished careers in several science fields. Looking out our office window, we frequently saw Sidney Chapman walking up the hill to his visitor's office. Many similarly famous visitors were in residence from time to time at HAO/CU. As graduate students, it was our job to run the slide projectors during lectures by visitors – a good way to learn about recent research.

While taking graduate courses at CU and working on HAO assigned tasks, the vigorous research environment in Boulder stimulated me to develop some of my earlier work at Lockheed and Mt. Wilson. One of these projects involved my interest in the roots of polar plumes observed at the 1963 total solar eclipse as mentioned earlier. Guided by Gordon Newkirk, improved analysis better supported the idea that the roots of coronal solar plumes did coincide with Ca II bright magnetic network features (Newkirk and Harvey, 1968). As part of this work, I learned about current-free modeling of the coronal magnetic field from photospheric measurements.

Hermann Schmidt was a visitor at HAO early in my time there. He had written a set of Fortran programs to extrapolate local surface magnetic field maps into the corona using a potential field approximation (Schmidt 1964). Having a number of digital magnetograms in my possession, it was natural to try his programs on some of these. He generously lent me his programs. I wrote a memo to myself about his programs in order to understand how they worked. Through a series of errors, my memo was published as HAO Astro-Geophysical Memorandum 172 without coordination with Schmidt. This was a serious mistake on my part for which I apologized to Hermann. An obvious spinoff from using Schmidt's programs was to try to extend the idea to extrapolating the entire solar surface magnetic field into the corona. I had access to enough daily Mt. Wilson maps of the line-of-sight magnetic field to build up a diachronic (over time) synoptic map covering one full solar rotation around the time of the 1966 total solar eclipse. With this as an initial condition, I experimented with various extrapolation methods and attracted interest from Gordon Newkirk and Martin Altschuler. Together we produced one of the first extrapolations of the full coronal magnetic field and found reasonable agreement with observed structures (Newkirk, Altschuler, and Harvey, 1968).

Before starting a PhD thesis a major academic hurdle for me was to pass the formidable AG Department Comprehensive Exam. By some miracle, on my second try, I passed the written exam with a score high enough to avoid an oral component. I then proposed in September 1966 to do a study of the magnetic fields in active region prominences taking advantage of my connections with Lockheed and Mt. Wilson, but based mainly on the unique HAO magnetograph installed on the 40-cm coronagraph at Climax. David Rust had used an analog version of the instrument for his doctoral thesis on quiescent prominences (Rust, 1966). One of my HAO tasks was to write a data acquisition and reduction program for the new Digital Equipment Corporation PDP-8 computer at Climax that replaced the original analog data recording system. I wrote a detailed flow chart using all the available space on three large classroom black boards. It was then relatively easy to convert it into assembly language instructions. A benefit of this was learning about proportional-derivative-integral servo loops in digital form. As part of my thesis project, I proposed to obtain some vector magnetograms at Kitt Peak National Observatory (KPNO) using the new McMath Solar Telescope. Somehow my convoluted involvements with Lockheed, Mt. Wilson, KPNO, and HAO produced enough data for a lengthy thesis. My thesis committee wanted even more material after reading the first draft with the result that my thesis was, at the time, the longest one ever produced in the AG department – a triumph of quantity over quality (Harvey, 1968, 1969).

I was not acquainted with the solar program at KPNO but was very warmly welcomed by the staff, especially by Bill and Dorothy Livingston, and took to Bill's equipment like a duck to water. Little did I know then that this would be the home of my future career. On one Frontier Airlines jet flight from Denver to Tucson to observe, I was the only passenger. Air travel is certainly evolved since then.

When I defended my doctoral thesis, a number of foreign visitors to HAO were in the audience. Perhaps they did not understand the nature of the defense but that turned out to be helpful to me. Frequently, when I was questioned about something, Ulrich Anzer enthusiastically answered before I could respond. My degree should be shared with him.

## 8. Life with Karen

My first encounters with Karen Angle were at UCLA where she enrolled in the astronomy department two years after me. Her father also worked briefly at Lockheed where he interacted with the solar observatory group. He had triggered Karen's interest in solar research, and she was one of the few people at the time to observe a white-light flare (Angle 1961). Her flare observation raised interest in what the Lockheed flare patrol showed. I prepared an illustration comparing the observations that she included in her paper. After that, we saw each other at UCLA and at Lockheed where she took on a part-time position working mainly with Sara Martin. Karen also worked as a Summer Research Assistant for Bill Livingston in the Solar Division of KPNO before I did any work at KPNO.

While working on her master's degree, Karen worked full time at Lockheed from 1966 supported by contracts that Sara had obtained, and she very much enjoyed working with Sara and Harry. One of the research projects (Harvey *et al.*, 1971) changed both of our lives by bringing us into frequent contact. Romance ensued (see Figure 5).

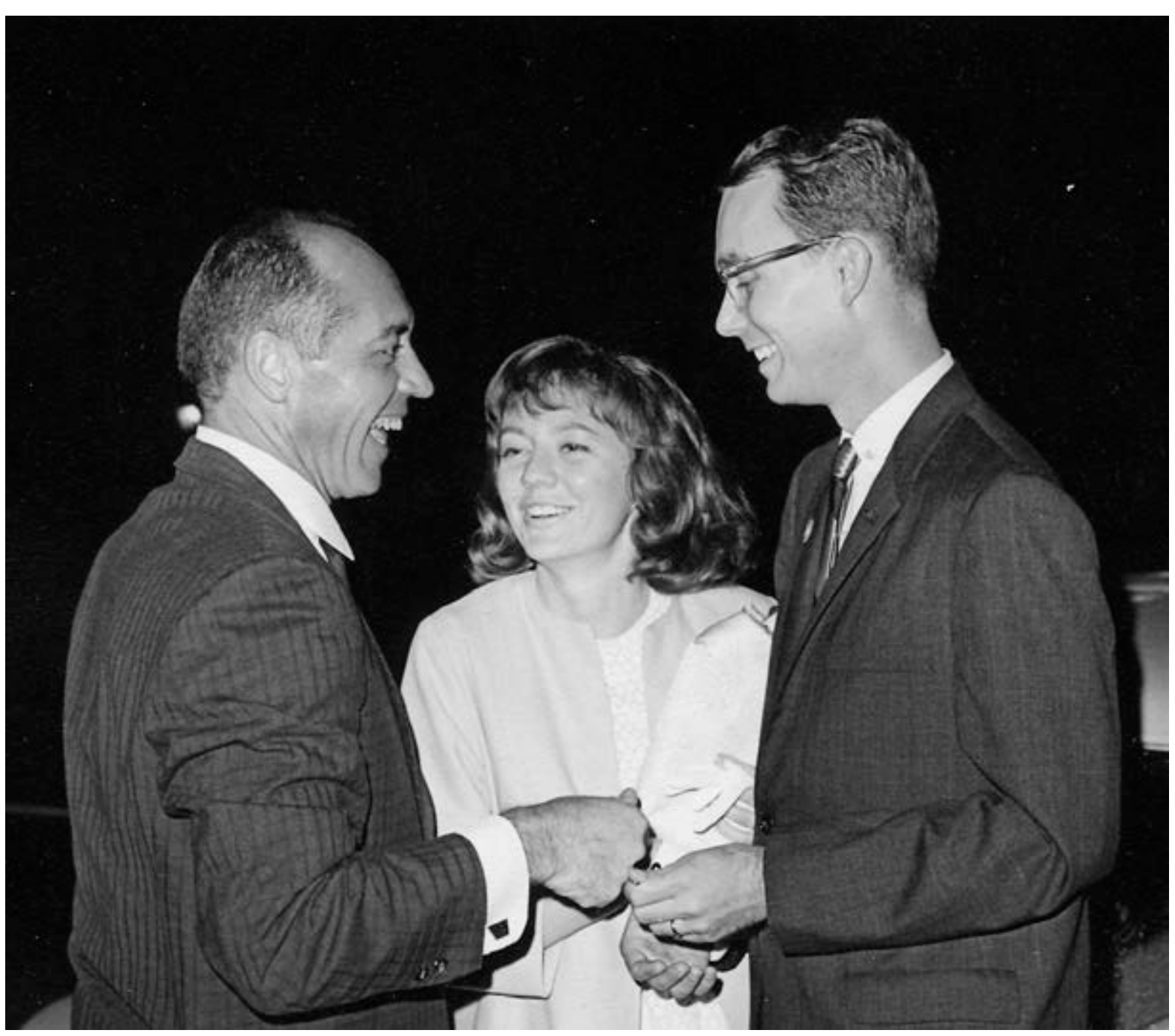

**Figure 5** Harry Ramsey (left) congratulating Karen and me at our wedding in October 1968.

After we married, we lived in Boulder for a few months as I finished my doctoral thesis. Karen was a superb typist and greatly helped me with drafts of my thesis. She was no longer affiliated with Lockheed, was briefly a visitor at HAO, and had finished her master's degree. With the end of my doctoral work, it was time to face our future. Since funding for science in 1969 was still generous, I was lucky to have several post-doctoral job offers including Caltech, Lockheed, Mt. Wilson, HAO, SPO, Aerospace Corporation, University of Hawaii, and KPNO. Karen and I both liked the people at KPNO and working with the largest solar telescope in the world was irresistible. So, even though offered the second lowest salary of the options, we decided to go to KPNO. (I later learned that KPNO salaries were invariably the lowest of a group of 25 peer organizations.) Unusually, I was hired before I actually finished and defended my thesis thanks to support from HAO's O. R. White that I only learned about much later.

Part of my accepting a position at KPNO was an agreement that Karen would be given long-term visitor status there to allow her to continue the work she was doing with Sara Martin at Lockheed and with me at KPNO. She became an integral part of the Solar Division of KPNO. When Lockheed's solar group moved to Palo Alto, Karen's Lockheed employment had ended, as did Sara's. Doug Martin had started an optics company (SpectraOptics) that incidentally provided a way for Sara and Karen to continue their solar research, and Karen became an employee. This continued until Sara obtained a position at Caltech in Hal Zirin's solar program. Karen then continued unfunded research work for a while.

Our son, David, was born in 1974, the same year I was promoted to a tenured position. (In those simpler times, I was not even aware that I had been proposed for tenure until KPNO Director

Leo Goldberg congratulated me one day, after the fact.) Karen wanted something to do at home while enjoying motherhood. Gordon Newkirk had been advocating the value of an index of articles for the *Solar Physics* journal and Karen agreed to do the indexing job. Karen had an observing run on Kitt Peak a few months after David was born, and so the three of us spent several days observing with the new vacuum telescope. (Much later, David was asked if he wanted to become a solar astronomer and said, "No, no, no! Two in the family is already enough." David is now a very successful mechanical engineer.)

Karen shared interests with many of the visitors to the KPNO solar division, and it was not long until she had started significant research about active region properties with Vic Gaizauskas, O.R. White, and Kees Zwaan. I also partnered in some of these projects, but usually at a low, observational support, level.

## 9. Solar Physics Research Corporation

In 1975, Karen submitted a proposal for a research project to NASA and it was accepted with one condition. The NASA solar program manager, Dave Bohlin, told her that the award could not be made to an individual. So Karen and I created a legal entity in the state of Arizona, the Solar Physics Research Corporation (SPRC), in order to accept the $4,000 grant. SPRC continued quietly with a series of small grants to Karen, and as a convenient conduit for NASA support of scientific meetings. I did the simple corporate paperwork separate from my regular KPNO job. One year, the National Science Foundation (NSF) randomly picked SPRC for an in-depth financial audit and sent a two-person team to conduct the week-long audit at our dining room table. We passed with two minor requests for bookkeeping changes. Simplicity ended when an NSF program officer asked if SPRC would accept transfer of a grant supporting Charles Lindsey from Hawaii to SPRC. An informal agreement with the now-renamed National Solar Observatory (NSO) gave Charlie long-term visitor status. A request for a similar arrangement for Doug Braun followed. Unplanned growth continued with SPRC eventually paying the salaries of Sara Martin, Hugh Hudson, John Jefferies, John Barentine, Ken Topka, Yeming Gu, Janos Bartus, and Aki Takeda, located in California, Japan, Colorado, and Arizona. NSO benefited at no cost by a significant increase in research activity and other in-kind contributions to the program.

This successful collaboration between NSO and SPRC eventually attracted the attention of the corporate lawyer of the Association of Universities for Research in Astronomy (AURA, the legal entity operating KPNO). In spite of strong arguments by both the NSO program management and SPRC, a forced separation occurred, weakening both organizations. SPRC rented its own office and paperwork became rather burdensome until most SPRC people found more permanent positions and their grants were transferred to other organizations. When Sara Martin briefly joined SPRC, a dispute with Hal Zirin (Martin, 2015) followed along with threats by Caltech of legal action against SPRC. Fortunately, reason prevailed and nothing came of this.

## 10. Work in the Solar Division of Kitt Peak National Observatory 1969-1983

Starting in 1969 as a post-doctoral, assistant-level member of the Solar Division of KPNO, I was eager to contribute to the exciting work being done with the new McMath telescope. The staff in the early 1970s included Keith Pierce, Bill Livingston, Jim Brault, Neil Sheeley, Bob Milkey, John Kirk, Don Hall, Jim Heasley, programmers Charles Slaughter and Richard Stevens, telescope manager and observer Bruce Gillespie, and technicians Richard Aikens and Mike Doe. I learned from all of them. My main job was to support visitors who came to use the McMath telescope by setting up and operating instruments with them. As a staff member I was also expected to compete for observing time to do my own or collaborative projects. It was an exciting time with many short- and long-term visitors from all over the world coming to use the facilities. Some of the longer-term and frequent international visitors included Arvind Bhatnagar, Ron Giovanelli, Eberhard Jensen, Oddbjorn Engvold, Y. Ohman, Peter Wilson, Edith Mueller, Kees Zwaan, Victor Gaizauskas, and Jan Stenflo. There were several other post-docs at KPNO and across the street at the Steward Observatory of the University of Arizona, and Karen and I socialized with many of them (including Don Hall, Bob Kirshner, Bob Milkey, Bob Carswell, and their wives).

### 10.1. ORGANIZATIONAL CHANGES AT KPNO

In the early 1970s, the new director of KPNO, Leo Goldberg, decided to replace the divisional structure of KPNO with a more centralized, programmatic organization. One benefit from this reorganization was improved interaction between solar and non-solar programs at KPNO. I served a fifteen-month term as chairman of the Solar System Program and I'm grateful to have learned early in my career that I was not cut out for scientific management! Funding for operations and new projects was generous until about 1975 when the first of more or less decadal budgetary crises occurred. That year the KPNO solar scientific staff reached a low count of four: Keith Pierce, Bill Livingston, Jim Brault, and me. The Solar System Program faded away when former Rocket Program and Planetary Program scientists and engineers left KPNO for other positions.

In 1976 the operation of Sacramento Peak Observatory (SPO) was transferred by the U.S. Air Force to NSF and Air Force funding was reduced. This additional impact on the NSF budget prompted efforts at KPNO to seek resources from other sources and caused somewhat strained cooperation between SPO and KPNO. All this management and budgetary turmoil disrupted on-going and proposed programs but also opened new opportunities. I continued to try to support and do solar research with the same enthusiasm as I had during my first years at KPNO and concentrated in a few specific areas described below.

### 10.2. COOPERATION WITH NASA

The Apollo lunar program was in full swing in the early 1970s and planning for the follow-on *Skylab* space station was well underway. Solar research was a major scientific component of *Skylab* and eleven of the NASA astronauts involved in *Skylab* visited Tucson for briefings and demonstrations of solar observing along with their solar mentor, Frank Orrall. Neil Sheeley and I enjoyed showing these famous men (including the third and fourth persons to land on the moon) how solar observing was done at Kitt Peak (Figure 6). My career-long friend Neil Sheeley recently published his recollection of our encounters with the astronauts along with other fascinating vignettes of his career (Sheeley, 2019). *Skylab* mobilized most of the US solar physics community, and a large meeting was held in Tucson to plan how best to support the *Skylab* solar observations from the ground. Two outcomes of the meeting were plans to make daily synoptic full-disk magnetograms using the McMath telescope during the *Skylab* mission and a proposal by Bill Livingston to build a solar Vacuum Telescope on Kitt Peak to make long-term dedicated synoptic observations. My parents, Karen, and I were thrilled to witness the awesome Saturn V launch of *Skylab*. During the mission I was called from *Skylab* (indirectly) by astronaut Owen Garriott who wanted to know more about Ellerman bombs. On a later occasion I happened to be seated next to him on a plane flight and was intrigued that he unfastened his seatbelt the moment the plane's wheels touched the ground. Trained for a quick emergency exit, I assumed.

Later in the 1970s I became involved in planning for solar observations with NASA's Space Shuttle and the *Spacelab 2* shuttle flight that featured several solar projects. Bill Livingston and I from KPNO were nominated by Alan Title as potential payload specialists to fly on the shuttle. A group of over a dozen candidates spent an interesting week undergoing tests in Houston. Subsequently, eight finalists were interviewed by a NASA committee in Huntsville, and I was offered a position as a backup payload specialist (Figure 6). Suffering at the time from a head cold, I declined the offer, later thinking "What did I just do?" But it turned out to be the right decision. Working with Alan's team in Houston supporting the *Spacelab 2* mission from the ground was an exciting and memorable experience.

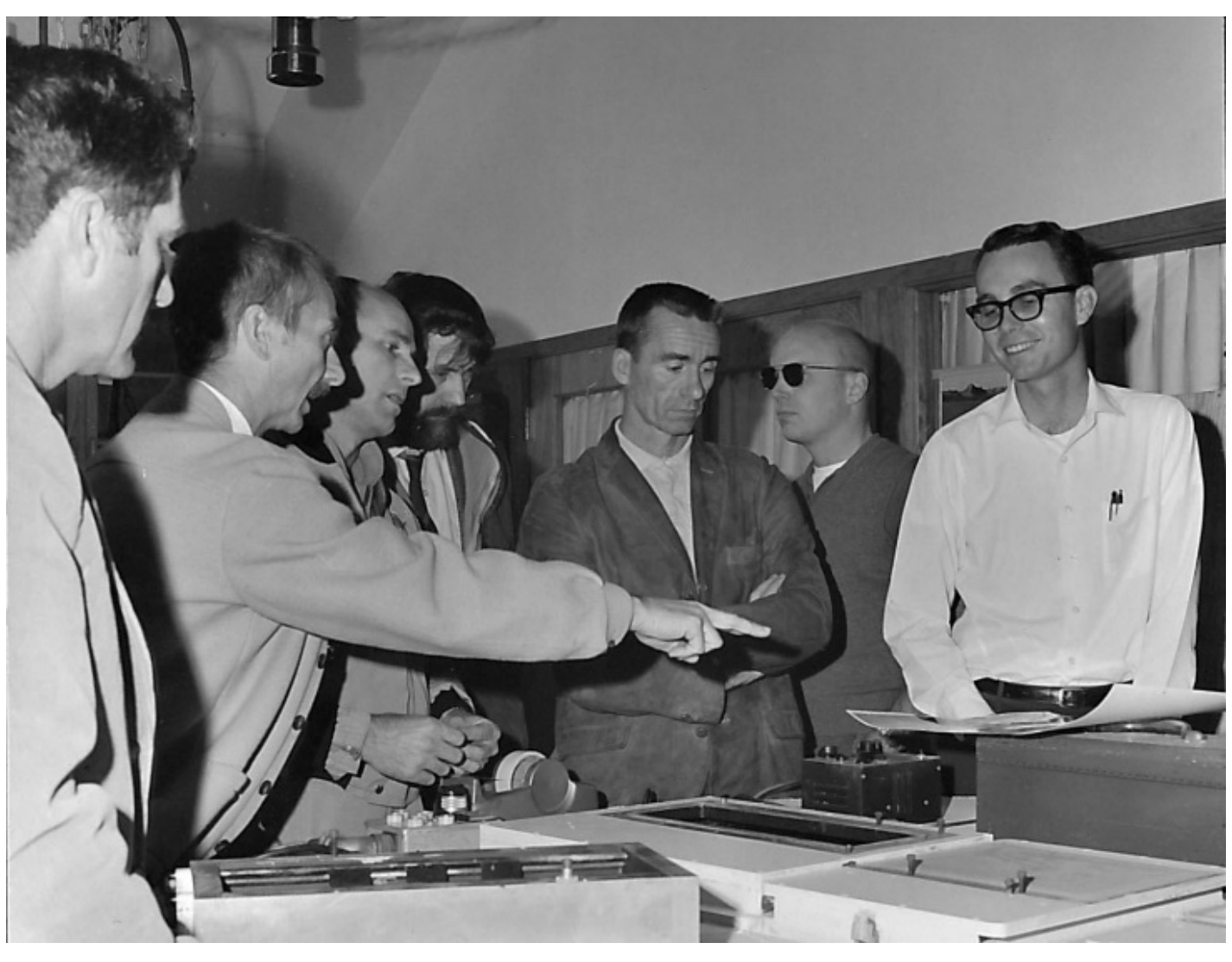

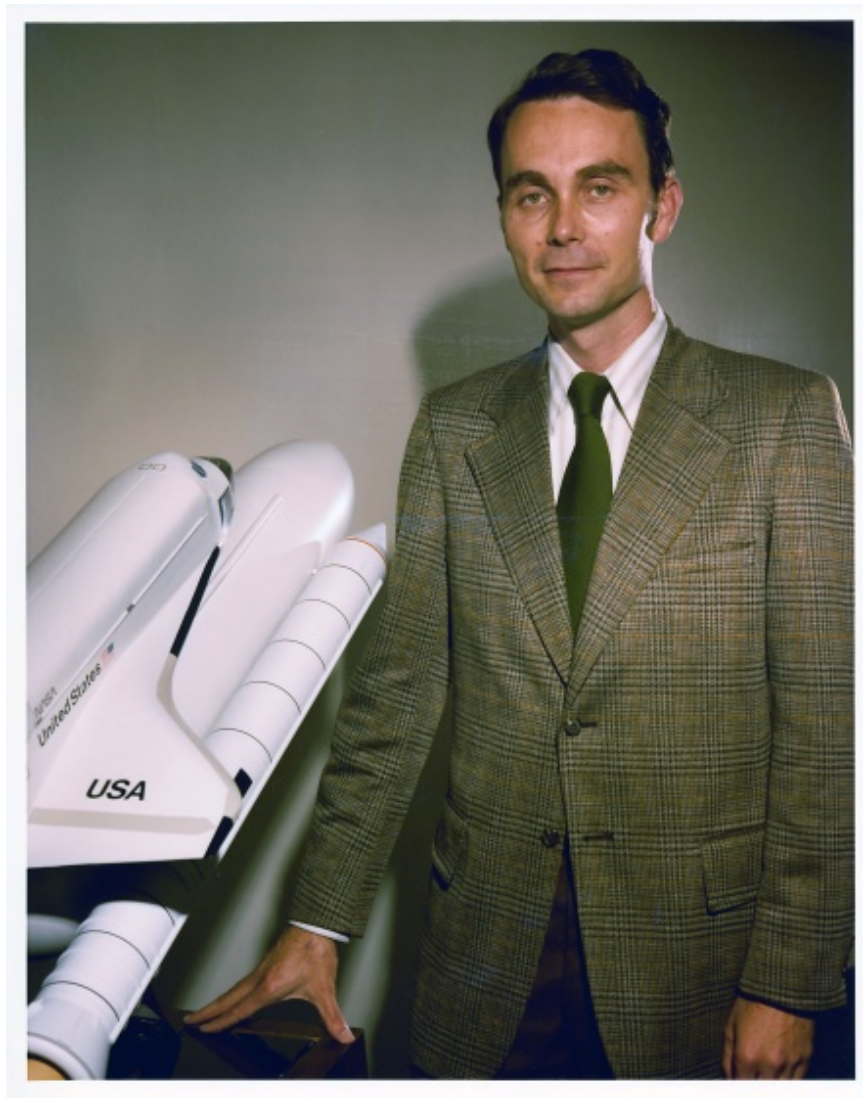

**Figure 6** (Left) *Skylab* astronaut candidates visited the McMath Telescope observing room in 1971 to learn about solar observations. (Left to right) Joe Kerwin, Owen Garriott, Alan Bean, Frank Orrall, Walter Cunningham, Story Musgrave, and me. (Right) My official 19 April 1978 NASA Payload Specialist Candidate photo.

Throughout my career I served on many NASA advisory groups and was a co-investigator on several proposals for space solar missions, but, until very recently, I never became deeply involved in space solar observations.

10.3. SPECTROPOLARIMETRY

The great strengths of the McMath solar telescope were its light gathering power over a wavelength range limited only by the atmosphere and a well-equipped spectrograph to deconstruct that light. These attributes were highly valued in those long-ago times since the principal detectors were single-pixel photoelectric devices that seldom had quantum efficiencies better than 10%, and ~1% efficiency photographic chemical emulsions. Guided by my generous principal mentor, Bill Livingston, I gravitated toward high-sensitivity spectropolarimetry. I had used his very versatile dual-channel vector magnetograph (Livingston and Harvey, 1971a) in my thesis work, and so my first work at KPNO utilized it for some projects with Bill and visitors. One puzzle of the day was why Babcock-type magnetograph measurements gave quite different results depending on the spectrum line used. I thought that different temperature sensitivities of the lines were the main cause (Harvey and Livingston, 1969). But evidence was building that the measurements were discrepant mainly because much of the magnetic flux was in the form of intrinsically strong fields that were measured incorrectly due to the non-linear instrumental response to Zeeman splitting (Howard and Stenflo, 1972). The latter view was, of course, correct, though controversy remained for many years (*e.g.* Zirin and Cameron, 2001). My mistaken idea about the discrepancies was one of my many missed opportunities.

Stimulated by Alan Title's frequent visits to use the McMath telescope with his spectra-spectroheliograph film recording system (Title and Andelin, 1971), I appreciated that measuring entire line profiles was far more valuable than just portions of them. So I built a photoelectric digital recording instrument that we called a "line-profile Stokesmeter" or a "profileograph" depending on whether polarimetry or spatial mapping was done. An unpublished example of the latter is shown in Figure 7. Similar unpublished maps near disk center showed that there were occasional strongly blue-shifted features in quiet disk areas in the chromosphere. Regrettably, I failed to publish these results and now these features are recognized as rapid blueshifted excursions (RBEs, Langangen *et al*., 2008).

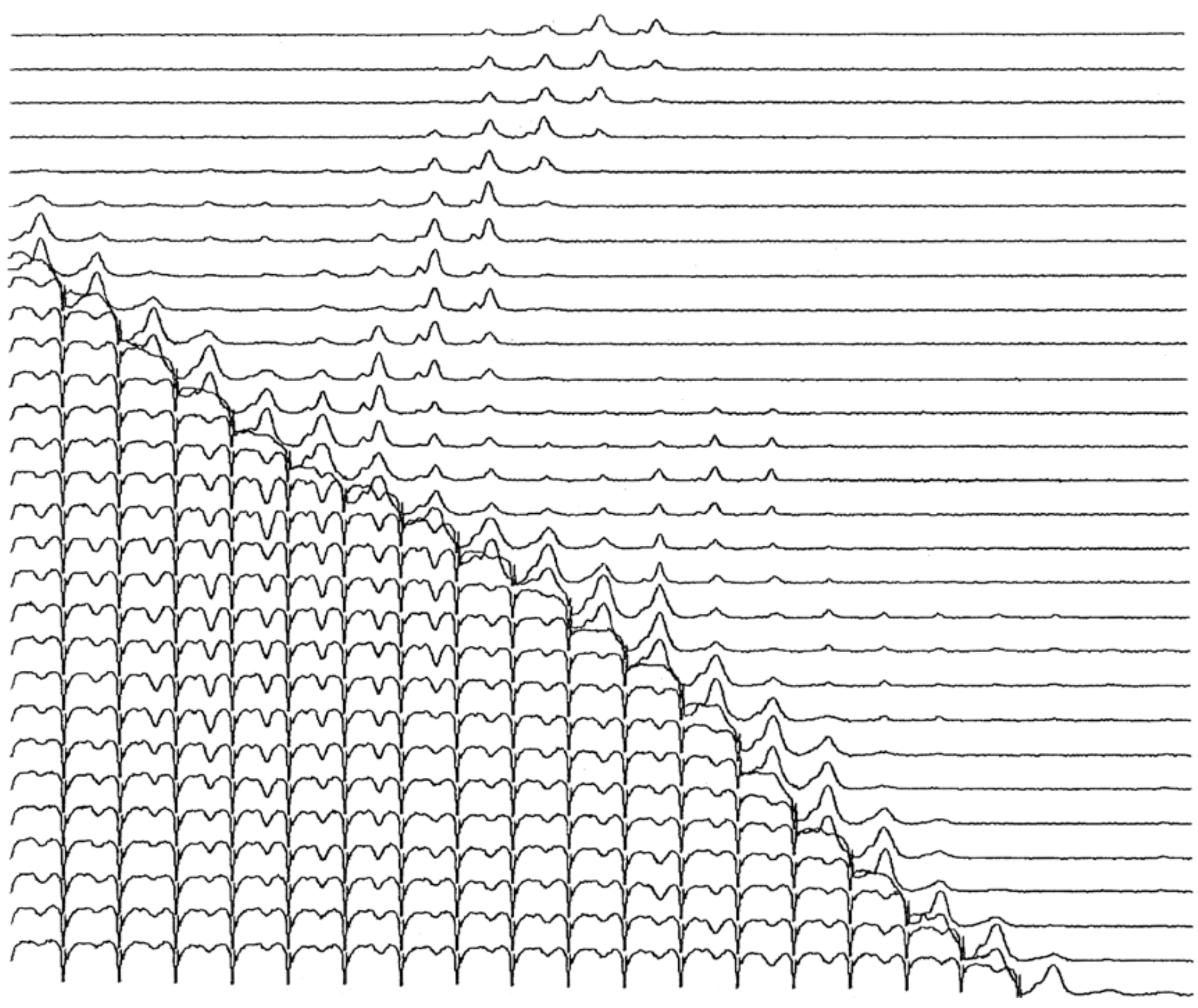

**Figure 7** Portion of a spatial mosaic of profiles of the HeI 10830 Å multiplet near the limb showing a prominence in emission and on-disk absorption features. Observation date 1973 September 1. The spatial steps are separated by 2.5 arcsec.

The principal result from the line profile instrument was the first numerical inversion of I and V Stokes spectral line profiles, which supported the kilogauss nature of photospheric network magnetic fields (Harvey, Livingston, and Slaughter, 1972). Don Hall had recently completed an infrared solar spectrograph at the McMath telescope. He noted seemingly strong Zeeman splitting of Fe I 15648 Å in the umbra of a sunspot. So it was natural to see if network features also showed strong Zeeman splitting. Using polarization optics, we found that they did, thus directly indicating their kilogauss field strength (Harvey and Hall, 1975). At that time, the g-value of the line was unknown, and so I set up an iron arc source in a kilogauss magnetic field and we obtained polarized spectra. The only result from that exercise was nasty sunburn of my face from the powerful ultraviolet emission of the arc. The infrared line turned out to be too faint to measure its Zeeman splitting.

Ron Giovanelli was a frequent visitor to KPNO and promoted the diagnostic power of the chromospheric He I 10830 Å multiplet. My long-standing interest in the chromospheric magnetic field, plus Don Hall's knowledge of infrared detectors, led to construction of an infrared detector system for Bill Livingston's versatile magnetograph. Equipped with this new detector, Don and I made the first magnetograms with the He I 10830 Å line (Harvey and Hall, 1971). Though crude by today's standards, the magnetograms revealed the diffuse and weaker nature of the chromospheric magnetic field compared to that in the photosphere.

Ferdinand Ellerman (Hale and Ellerman, 1920) constructed a photographic map of the sunspot spectrum. After 50 years his map was still in use, and so it seemed time to make a more modern version using the powerful spectrograph at the McMath telescope. To analyze polarization in front of the spectrograph, I constructed a large Babinet compensator, using quartz crystals given to me years earlier by Lockheed's George Carroll. I built a composite filter so that both the sunspot and surrounding photosphere could be captured in one single photographic exposure. I programmed the computer-driven spectrograph and 70 mm film camera to take a series of about 200 spectra with the right exposure times for the spectral region from about 3800 to 9200 Å. Using hypersensitized emulsions, the atlas was extended into the infrared. The atlas was published as a series of photographic prints and is now available more usefully as on-line scans (ftp://nispdata.nso.edu/pub/polatlas/). This atlas project led to findings of anomalous Zeeman splitting in molecular lines (Harvey, 1972a, 1973) and tables of diatomic molecular lines (Boyer, Sotirovski, and Harvey, 1975, 1976, 1978, 1982).

With the advent of Jim Brault's magnificent Fourier transform spectrometer (FTS) at the McMath telescope and its superlative capability to capture large wavelength regions with excellent spectral resolution, I worked with Jim to develop a polarimetric capability, especially in the infrared (Brault, 1978; Harvey *et al*., 1980). This included developing a mechanism to compensate for the significant polarization of the McMath telescope (Harvey, 1985). The FTS polarimeter project was stimulated by several visits to Kitt Peak by Jan Stenflo. I very much enjoyed working with Jan and supporting his observing runs, and we became good friends. The FTS used silicon diode detectors that had poor sensitivity at short wavelengths. So for this short wavelength spectral region we combined parts of Bill's dual-channel vector magnetograph with the photoelectric spectrum scanner of the large spectrograph. This allowed a much larger spectral region to be scanned compared to my earlier profile scanner. Figure 8 is a snapshot of a connection diagram of this setup extracted from my observing notebook. With this equipment, Jan confirmed and extended his earlier work on anomalous linear polarization associated with certain spectrum lines observed near the solar limb (Stenflo, Twerenbold, and Harvey, 1983).

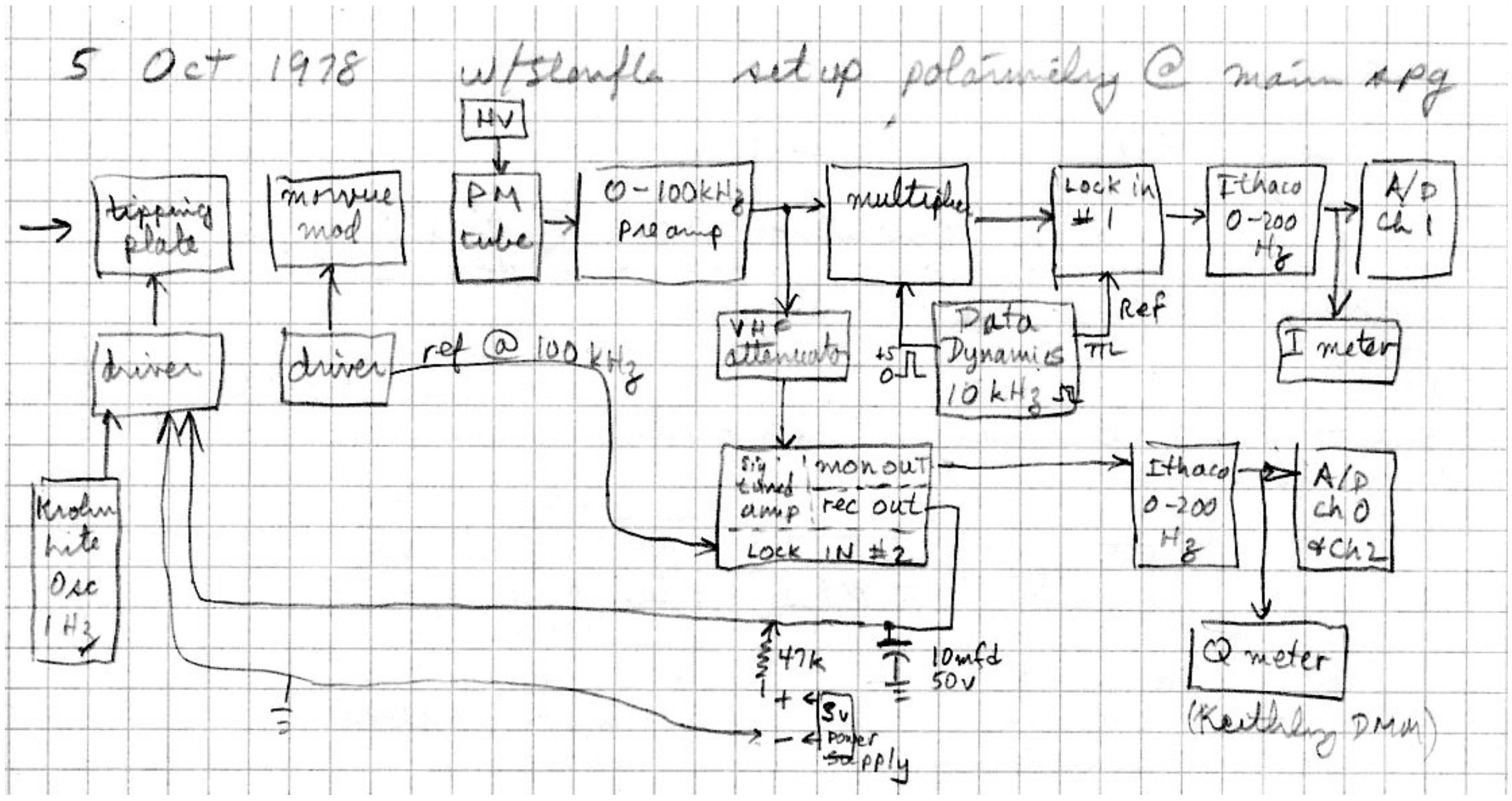

**Figure 8** Typical 1970s combination of pieces of existing equipment arranged to do I and Q Stokes polarimetry in the near ultraviolet using the McMath telescope and grating spectrograph.

The FTS polarimeter worked best at longer visible and infrared wavelengths, and quiet sun linear polarization effects were generally below the noise level. Accordingly, circular polarization observations of magnetic network and active regions were the main targets for the FTS (*e.g.*, Stenflo *et al*., 1984). Single observations could capture the Stokes I, Q, and V spectra over spectral regions limited mainly by the atmospheric wavelength dispersion blurring of the target feature across the observing aperture. The instrument was successful, but reduction of the interferograms was quite complicated due to squeezing three separate channels of data into a single recorded range. With the addition of infra-red detectors, it later became possible to extend the FTS polarimeter wavelength limit to the transmission cutoff of the fused silica and calcite polarizing optics near 2.8 $\mu$m (Rüedi *et al*., 1995).

10.4. SPORADIC SYNOPTIC OBSERVATIONS

Full-disk maps of the solar magnetic field were hard to make with the Livingston dual-channel magnetograph. So, in the late 1960s, Bill designed a 40 spatial channel magnetograph to enable more efficient mapping of the field (Livingston, Harvey, and Slaughter, 1971). Construction was underway when I arrived, and Bill invited me to participate in completing the system. My role was computerized calibration, reduction, and display of the expected data. (Later I learned that Bill suggested my appointment to KPNO because I had useful computer experience.) The instrument was successful with its first results providing daily magnetograph observations around the time of the March 7, 1970 total solar eclipse (Livingston, Harvey, and Slaughter, 1970). Occasional full-disk and local area scan maps were made until support of the NASA

*Skylab* mission became a high priority in 1973. For the duration of the *Skylab* mission, more or less daily observations were interspersed with more traditional projects at the McMath telescope. Among the early results was an observation of the inner-network magnetic field (Livingston and Harvey, 1971b).

## 10.5. SUSTAINED SYNOPTIC OBSERVATIONS WITH THE KITT PEAK VACUUM TELESCOPE

Leo Goldberg was a strong advocate of solar observations from space and, as director of KPNO, he assigned high priority to a Livingston proposal to build a new telescope and magnetograph system to support the 1973-74 *Skylab* mission. The 60 cm aperture vacuum telescope on Kitt Peak (KPVT) was designed and built in record time thanks to project engineer Dale Schrage (Livingston, *et al*., 1976a).

My roll was the computerized observing program. As with the HAO magnetograph, I prepared a detailed flow chart on large sheets of paper and then worked closely with programmer Charles Slaughter to convert it into assembly language code. We operated both the older 40-channel at the McMath telescope and a new 512-channel magnetograph at the KPVT until 1975. After the *Skylab* mission, funding was much reduced and we built on *Skylab* partnerships with NASA and the National Oceanic and Atmospheric Administration (NOAA) to continue the daily synoptic observations made with the KPVT. These mostly informal arrangements led to much-welcomed additions to the KPNO solar staff. Harrison Jones joined from NASA, as did Thomas Duvall, Jr. after he finished a post-doc appointment with the KPNO Solar Program, and NOAA employee Frank Recely was stationed at Kitt Peak and became the chief observer until his retirement. These people enabled the synoptic magnetogram program at KPNO.

The Reticon™ detectors used in the KPVT 512-channel magnetograph were sensitive at the 10830 Å He I multiplet (Livingston *et al*., 1976b). This capability enabled full-disk observations that showed the locations of coronal holes. Coronal holes were much studied at that time and we were able to continue such synoptic observations past the end of the *Skylab* mission. Among the results were charts showing strong, long-term association between geomagnetic activity and coronal holes (Sheeley, Harvey, Feldman, 1976; Sheeley and Harvey, 1978, 1981). Another important discovery was on-disk confirmation of the occurrence of bipolar, transient coronal holes (now called coronal dimmings) associated with strong flares (Harvey and Recely, 1984).

## 10.6. ECLIPSES

Total solar eclipses offer opportunities to study the Sun in ways not normally available. The KPNO solar program frequently used this opportunity through community programs organized by the National Science Foundation. Soon after I joined KPNO, preparations were started to observe the March 7, 1970 eclipse in southern Mexico. Bill Livingston and I thought it would be worthwhile to try to measure Doppler shifts of the Fe XIV 5303 Å line in order to measure the coronal rotation and local flows. With Mike Doe, we built a portable spectrograph that used an image intensifying tube and 70 mm film as the detector. A neon emission lamp provided simultaneous wavelength calibration. To improve spatial coverage, we used five entrance slits, knowing that the overlapping spectra could be separated later. Observations were successful (Little and Robinson, 1980). While Mike and I operated the spectrograph in Mexico, Bill stayed in Tucson and made the daily magnetograms from which our first diachronic (over time) synoptic map was constructed (Livingston, Harvey, and Slaughter, 1970). We did not know at the time that hundreds more synoptic maps would be in our futures. (The first solar synoptic map of which I am aware was constructed by Peters (1856) from his 1845-1846 sunspot observations.)

The eclipse spectrograph was deployed to Kenya in 1973, in 1980 to India, and finally, in 1983 to Indonesia (where clouds precluded good data). The results are summarized in Livingston *et al.* (1980) and Livingston and Harvey (1982). On the other side of the African continent in Mauritania in 1973, I attempted to make Doppler images of the corona using a purpose-built birefringent filter but a thick dust storm at eclipse time ruined that attempt. In one of my few excursions into nighttime astronomy, the eclipse spectrograph, equipped with an early CID solid state sensor, was used at the 4-m Mayall telescope to image the potassium emission cloud around α Ori (Lynds, Harvey, and Goldberg, 1977).

## 10.7. DOPPLER VELOCITY MEASUREMENTS

There was a controversy about short-period (seconds) Doppler oscillations in the early 1970s. Were they solar or some sort of seeing noise? My feeling was that they were caused by seeing-induced image motion moving spatial gradients of the Doppler velocity across an observed region of the Sun. I attended a concert by avant-garde composer John Cage and, while watching all the mechanical actuators he attached to various objects, I realized that using x-y drive signals to move an observing aperture in a small circular pattern at high frequency would allow one to observe local Doppler gradients at the same time as the standard Doppler signal (Harvey, 1970). This helped to prove the non-solar origin of short-period oscillations. I always regretted not crediting John Cage in my paper. Further observations at five magnetograms per second using the new 40-channel magnetograph settled the question (Harvey and Howard, 1972).

The Livingston dual-channel magnetograph only observed one spatial pixel at a time but was well equipped with multiple sensors in its spectrograph focal plane. This made it ideal for studying Doppler oscillations. Working with Arvind Bhatnagar and Ron Giovanelli, I contrived a way to make real-time Doppler shift observations of magnetic features uncontaminated by surrounding scattered light. The technique measured separate zero crossings of the $dI/d\lambda$ and V Stokes line profiles simultaneously (Bhatnagar, Livingston, and Harvey, 1972; Giovanelli, Livingston, and Harvey, 1978; Giovanelli, Harvey, and Livingston, 1978). Karen and I also used this instrument in an area scanning mode to study Hα Doppler shifts in an active region with many flares and found large-scale organization of Doppler shifts consistent with major shear along the polarity inversion line (Harvey and Harvey, 1976).

## 10.8. HELIOSEISMOLOGY AT KPNO

I had spent time at Mt. Wilson in 1963 observing for Bob Howard's study of the 5-minute oscillation but did not pay much attention to progress in the field. It was only after joining KPNO that my interest slowly developed. I've written about the history of helioseismology observations at KPNO and South Pole and refer the reader to that paper (Harvey, 2013). Here I give a more personal perspective of a few helioseismology activities in which I was involved.

John Leibacher and I first met in 1972 at the second meeting of the Solar Physics Division of the AAS shortly after his PhD thesis work. He was eager to confirm his theoretical results that the oscillations were acoustic waves trapped beneath the photosphere (Leibacher and Stein, 1971). He proposed making spatial-temporal power spectra of the five-minute photospheric oscillations using the Livingston dual-channel magnetograph in an area scanning mode. I was enthusiastic but did not have a good understanding of the type of scans needed. Joined by O. R. White, John and I attempted several observing runs, but due to instrument, data recording, weather problems, and my imperfect observing setups, none of these was successful.

After that disappointment my interest in oscillations waned until I assisted Eric Fossat with his observing runs at KPNO. He infected me with his great enthusiasm and became a friend as we visited each other's homes and families. At the time there was evidence of a 180-min solar oscillation, which if confirmed would be of great importance. We suspected that it might not be solar in origin and made observations that supported that supposition (Fossat, Harvey, Hausman, and Slaughter, 1977; Fossat, Grec, and Harvey, 1981).

Solid-state detector arrays were becoming available in the mid-1970s, and I proposed to construct an interface between the main spectrograph of the McMath telescope and one of the new detectors with the aim of a modern, general-purpose Doppler shift measurement capability for the users. Tom Duvall joined the KPNO solar group at about this time when Deubner's (1975) spectacular spectra of the five-minute oscillation power were announced. Tom and I decided to replicate and possibly extend Deubner's work with the new equipment. We concentrated on spatial scales between a solar diameter and about 1/100 of the diameter, used various optical methods to average the full solar disk into the spectrograph slit, and obtained a number of exciting results as described in my summary paper (Harvey, 2013). We also worked with Ed Rhodes, Jr. on smaller spatial scale observations. Unlike my earlier, failed observations, the results were very satisfying. Best of all were numerous interactions with other observers, theoreticians, and modelers. It was very exciting to be part of the early stages of helioseismology, which to me was a perfect example of how science should be done as a community enterprise. I am truly thankful especially to Tom Duvall, Douglas Gough, Stuart Jefferies, Ed Rhodes, and many others for allowing me the privilege of working and learning with them.

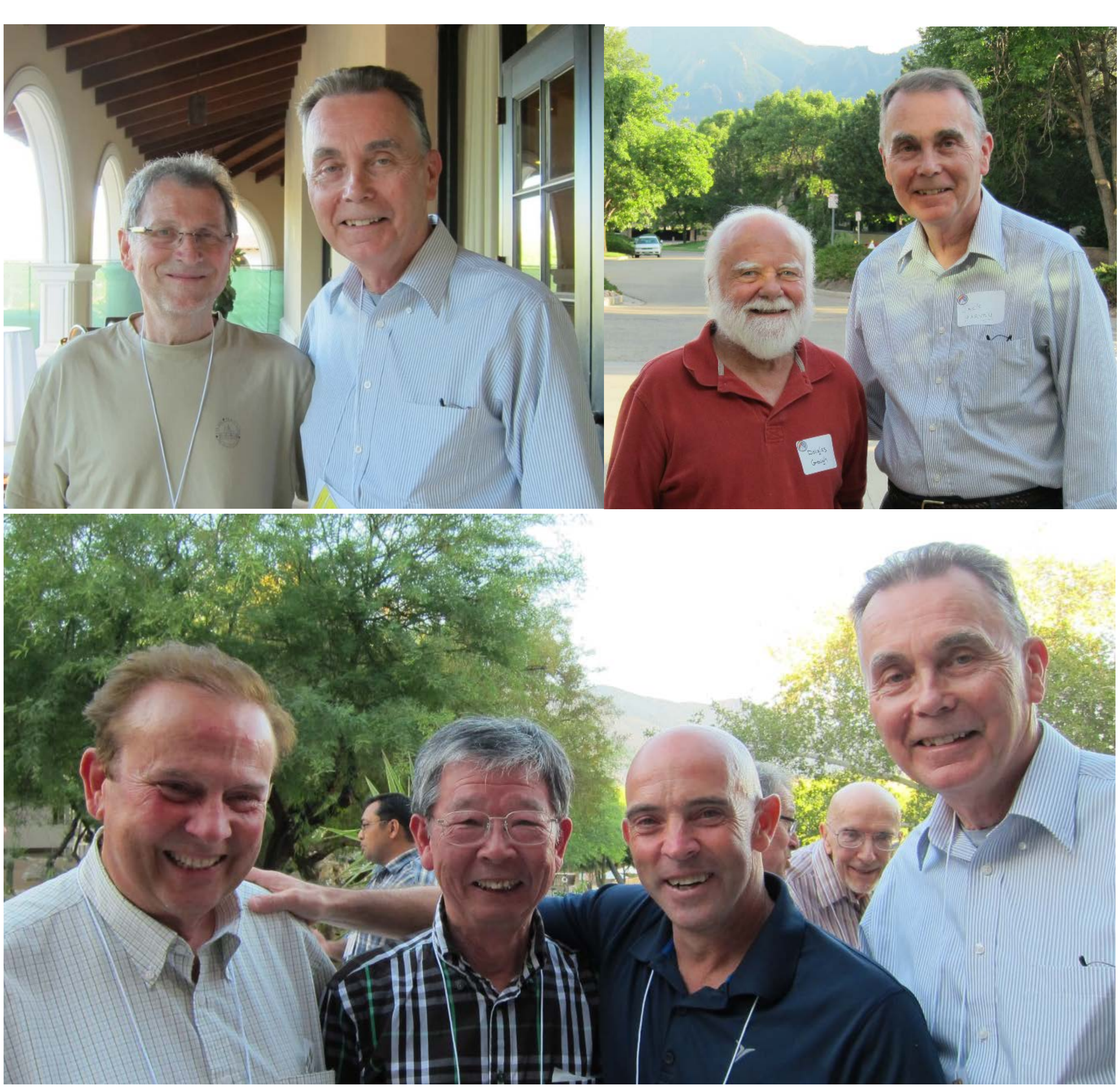

**Figure 9** Several of the giants of helioseismology who had major influences on me. Upper Left: With Eric Fossat in May 2013. Upper Right: With Douglas Gough in September 2015. Lower: With Tom Duvall, Yoji Osaki, and Stuart Jefferies (with my mentor Bob Howard behind my right shoulder) in May 2013.

The field of helioseismology was developing rapidly and after Eric's spectacular whole sun results from South Pole were announced (Grec, Fossat, and Pomerantz, 1980), Tom and I approached Martin Pomerantz at a meeting to see if there might be interest in attempting resolved disk helioseismology observations at South Pole. Martin was enthusiastic and fully supportive. We assembled a very simple, low-cost imaging instrument with mostly borrowed equipment. Tom wrote excellent data collection software and after one hour of successful testing in Tucson, we packed the equipment and were soon observing at South Pole. The observations were successful and led to four additional observing seasons with increasingly better equipment (Harvey, 2014). Stuart Jefferies soon joined the group and greatly contributed to our successes, and he continues South Pole observing today. My involvement in helioseismology radically changed with the start of the GONG project described in Section 11.2.

## 10.9. SPATIAL INTERFEROMETRY

My last formal academic course was solar radio astronomy taught by Jim Warwick. It included a lot about interferometry. Shortly after I arrived at KPNO, Richard Miller from the University of Chicago was visiting and gave several lectures on stellar interferometry that intrigued me. Visitors from France, Pierre Turon and Pierre Lena, also led informal discussions of interferometry that helped stimulate Jim Brault to build his FTS and, in me, an interest in spatial interferometry. Labeyrie (1970) had recently shown how to break through the seeing barrier with his technique of speckle interferometry. Using a modification of this method with the Mayall 4-m telescope, Roger Lynds, S. “Pete” Worden, and I made the first crude image of α Ori (Lynds, Worden, and Harvey, 1976). There was interest to extend speckle interferometry to solar features using the 1.5-m McMath telescope but various efforts at first (Harvey and Breckinridge, 1973), (Harvey and Schwarzschild, 1975), (Aime, Ricort, and Harvey, 1978) did not produce especially useful results due to detector deficiencies of that era. However, using the primitive 2-D McMath CID solid-state detector, Stachnik, Nisenson, and Noyes (1983) succeeded in making the first solar speckle image reconstructions. Today, speckle imaging and adaptive optics have revolutionized observational solar research.

One free day at the McMath telescope, I tried a really simple observation using wavefront division interferometry similar to that in wide use in radio astronomy. Here a virtual mask with two openings was placed on the telescope aperture and a fringe pattern would form in the focal plane if there were structures with a size corresponding to the spatial resolution provided by the spacing of the two apertures. With a suitable color filter and a fast exposure, I was able to photograph interference fringes with a simple 35-mm film camera (Harvey, 1972b, 1985; see Figure 10). Although personally very satisfying, the method produced only the unsurprising result that features in sunspots were at least as small as 0.2 arc sec.

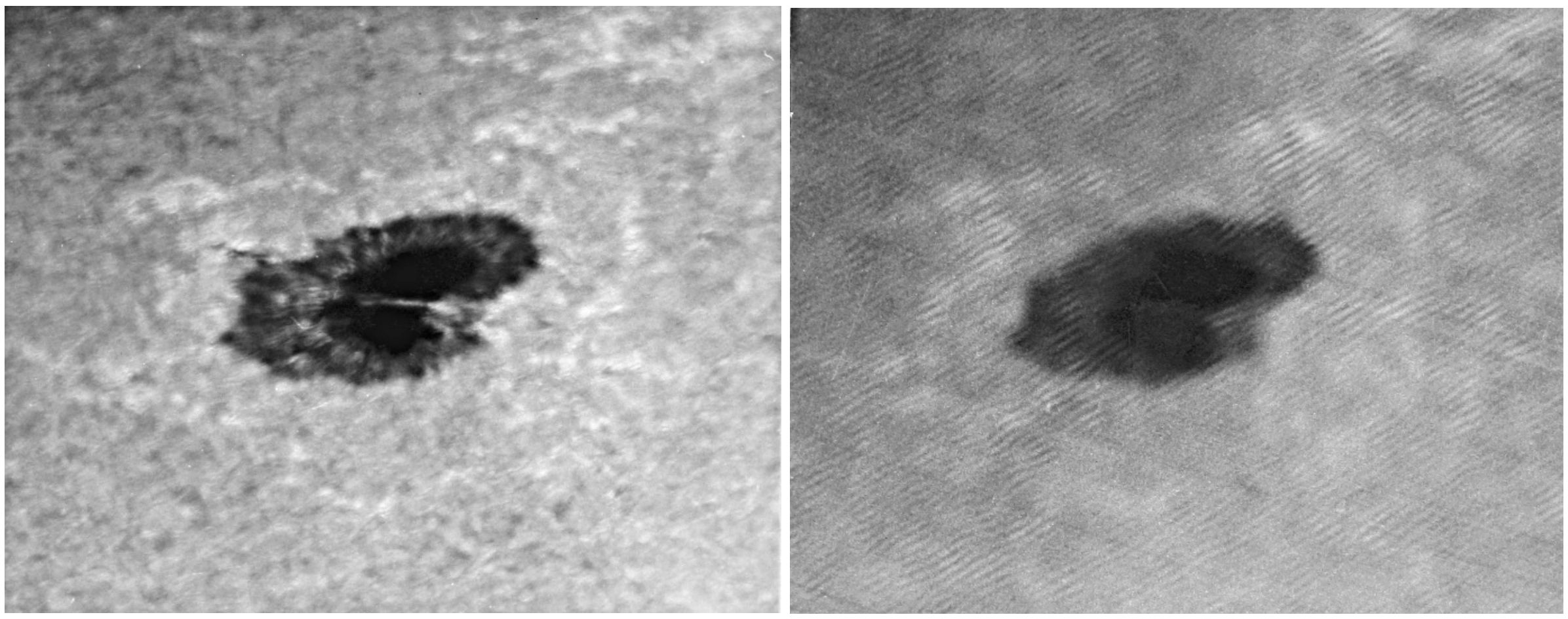

**Figure 10** Demonstration of solar wave-front division interferometry. Left: Normal photograph of a sunspot and surroundings. Right: Photograph showing two-beam interference fringes that indirectly indicate the presence of small spatial scale structures.

10.10. COMMUNITY SERVICE

As a tenured member of the staff of KPNO, I was expected to spend time serving the community in ways other than supporting users, doing original research, or refereeing papers and proposals. Beginning in the KPNO era, as noted in this section, I was an easy target for some of these activities. A rewarding aspect of committee service is meeting and interacting with new people interested in a wide variety of subjects and (usually) full of good ideas.

10.11.1. Journals

I became a member of the editorial board of *Solar Physics* in 1975, perhaps because I was one of the charter subscribers to the journal when I was a student. This turned out to be the start of a career-long involvement with the journal as described later. I also served on the editorial committee of *Annual Reviews of Astronomy and Astrophysics* for four years. This experience was very interesting because of the broad range of topics considered and subtle interplays of politics and personalities on the committee and with invited authors.

10.11.2. NASA groups

Starting with a 1975 Skylab workshop on coronal holes, I served on several NASA working groups. Some dealt with efforts to fly a large solar optical telescope on a space shuttle mission and later a large orbiting solar laboratory. These telescope working groups faced uphill battles that were ultimately lost for lack of support at high levels in NASA. However, one working group that was very successful, ultimately leading to the helioseismology projects SOHO/MDI and GONG, was the 1983-4 Science Working Group on the Measurement of Solar Oscillations from Space.

Another interesting experience was service on the first (1997) so-called Senior Review of NASA's solar and heliospheric missions. The goal was to assess the science return of NASA's operating missions to see if their costs could be reduced. NASA had set some severe budget targets but had not refined the Senior Review rules at this early stage. So Chairman Len Fisk astutely avoided the budget targets by clever strategies. It was a wonder to behold, as was the amazingly effective advocacy of Janet Luhmann for certain projects. Later senior reviews had to operate under more stringent rules. From 1988 through 1995, I served on one of a long series of science working groups developing what ultimately became the Parker Solar Probe. It is very satisfying to see the great science results from that mission, and to have witnessed the exciting launch of the mission.

10.11.3. American Astronomical Society Solar Physics Division (SPD)

I was fortunate as a student at HAO/CU to be able to attend joint meetings between HAO and SPO held annually during the 1960s in Santa Fe, New Mexico. The meetings were famously informal and very enlightening. While I was describing some of my thesis work, Charlie Hyder was unhappy about something that I said and gently, but surprisingly, threw a blackboard eraser at me. As detailed by Thomas (1999), these Santa Fe meetings stimulated the formation of the Solar Physics Division of the American Astronomical Society. In 1979, Karen was elected as treasurer, a position she held until 1995, and I was elected to vice chairperson and subsequently served as chair and then vice chair again with terms ending in 1982. The SPD survived this family takeover and celebrates its 50th anniversary in 2020.

10.11.4. National Research Council (NRC) and National Academy of Science

Serving on the 1980 Solar Physics Working Group of the third Astronomy Decadal Survey was my first introduction to competitive big science and an opportunity to help promote solar research. In this and 1990 and 2010 decadal survey activities I learned the power of compelling science and how a few articulate, and eloquent speakers can change the path of science. NRC has advisory committees of various kinds and serving on the 1981-1984 Committee on Solar and Space Physics was particularly exciting under the dynamic leadership of Louis Lanzerotti who showed me a lot about science politics. It is my experience that these various NRC committees only function well because of the dedicated staff of the NRC. I particularly thank Art Charo for his decades-long service to the science community.

10.11.5. National Science Foundation (NSF)

In 1982 I started a three-year term on the NSF Astronomy Advisory Committee. It was very pleasing to be on a committee that was exceptionally free of partisanship … until a certain new member joined. I learned that one partisan bad actor can spoil the whole committee experience and its work. It has been my experience, imbedded in the national observatory, that the small community of solar researchers seems to be much more united than does the much larger one of our nighttime colleagues.

## 11. National Solar Observatory under NOAO 1984-2010

The SPO and KPNO solar programs were combined as the National Solar Observatory (NSO) in 1984 as part of the newly-created National Optical Astronomy Observatory (NOAO). I was transferred from KPNO to NSO as part of this reorganization. My mentor from the 1960s, Bob Howard, became the first director of NSO and served until 1989. NSO budget conditions were constrained by NOAO management so there was a major program review and resetting of projects and resources to better support the mission of NSO. A consequence was longer term strategic planning and less ability to react to short-term ideas of visitors and staff. Unfortunately, the merger was not without friction between the former SPO and KPNO staffs.

Since the spectacular successes of various space solar missions reduced visitor interest in using NSO facilities for stand-alone research projects there was an opportunity to do collaborative projects using both ground and space-based facilities involving both former SPO and KPNO staffs. There was gradually developing community sentiment to either modernize the aging NSO facilities or to close them. With less pressure to support short-term visitor projects, long-term projects for the benefit of the community became a focus of NSO. Chief among these are the Daniel K. Inyoue Solar Telescope (DKIST), the Synoptic Optical Long-term Investigations of the Sun (SOLIS), and the Global Solar Oscillations Network (GONG). As a result of these trends I gave up the fun and excitement of short-term projects to concentrate mainly on improving the KPVT (and eventually replacing it with SOLIS), and developing the GONG instrumentation following our KPNO and South Pole helioseismology efforts. During the development of long-term projects, the NSO research staff sought and received short-term grant support from various government agencies. This support was very important in sustaining the vigor of NSO internal research activity. I did not get involved in the DKIST except as a member of a critical design review of one of the focal plane instruments. There was a large amount of luck and good timing in getting support for the three major projects. A major lesson is that it is highly recommended to have a project ready to start in case a funding opportunity arises.

### 11.1. MODERNIZING THE KPVT

When NSO was formed the KPVT program of daily full-disk magnetograms and helium spectroheliograms was a decade old and had proven its value both scientifically and for helping to forecast solar activity. The program was sustained by a tripod of support from NSF, NOAA, and NASA. When one leg was having budget problems the other two argued for continued support. Fortunately, the budget problems were never synchronized so the tripod never collapsed.

The original 512-channel magnetograph suffered from various electronic issues. These were addressed in a series of upgrades. But after nearly twenty years of use, it was time for a more robust and capable replacement. Instead of using just two linear detector arrays, thanks to the advance of detector technology, it was very attractive to use a CCD array to observe the full line profile to get more accurate Zeeman effect measurements. As usual, NSO was suffering budget shortfalls, and so NSO alone could not afford to do this project. A joint project between NASA and NSO was proposed with Harrison Jones as leader. The resulting focal plane instrument was called a spectromagnetograph (SPMG) and continued to operate into 2003 (Jones *et al*., 1992). In 1996, we started daily full–disk chromospheric LOS magnetogams using the 854.2 nm Ca II line and these provided many interesting new results (*e.g.* Harvey *et al.,* 1999). My role in the SPMG was relatively small because of my involvement with GONG and SOLIS.

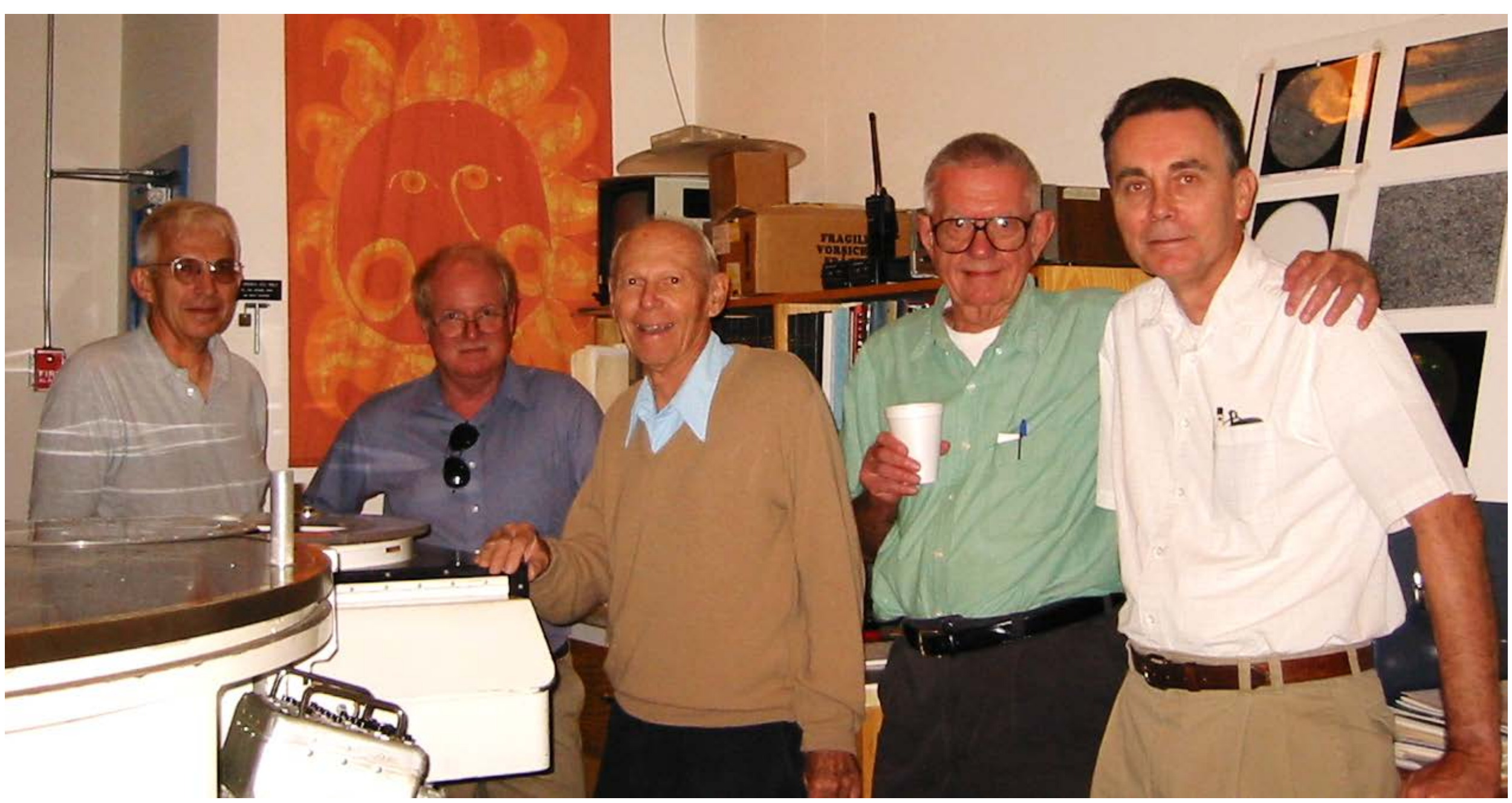

**Figure 11** Final day of observations at the Kitt Peak Vacuum Telescope with the NASA-NSO spectromagnetograph, September 21, 2003. (Left to right) Harrison Jones (leader of the spectromagnetograph project), Bruce Gillespie (long-time chief observer at the McMath-Pierce and Vacuum Telescopes), my mentors A. Keith Pierce and Bill Livingston (the founders of the McMath-Pierce and Vacuum Telescopes, respectively) and me.

The SPMG was replaced by a new instrument (see Section 11.3) in 2003. Figure 11 was taken on the last day of observations, which was almost exactly 30 years after the first KPVT observations. I wrote a brief history of the KPVT (Harvey, 2003) after its final observation. The KPVT building was demolished in 2019 but leaves a legacy of thousands of daily maps of the LOS magnetic field in the photosphere and chromosphere, and helium spectroheliograms. Nearly 2000 research papers, theses, and other publications have used KPVT data products, frequently in the form of diachronic and synchronic synoptic maps. KPVT archival data continue to be used. Data from the KPVT provided a number of firsts and discoveries. To name a few: measurements of the rotation rate of supergranulation; demonstration that most coronal holes rotate differentially;  association of open magnetic fields with coronal holes and high-speed solar wind throughout the solar cycle; on-disk transient openings of the coronal magnetic field to interplanetary space (now called coronal dimmings); perturbation of the frequencies of solar oscillations due to convective pressure;  and close correlation of the frequencies of global solar oscillations with the total surface magnetic flux. The KPVT was a success.

## 11.2. GLOBAL OSCILLATION NETWORK GROUP (GONG)

Just prior to the 1984 start of NSO, acting director Jack Zirker organized a joint meeting of the Kitt Peak and Sac Peak staff astronomers to consider new solar projects. One of these, stimulated by the burgeoning field of helioseismology and suggested by John Leibacher, was a global network of six identical instruments to observe solar oscillations for a period of three years, *i.e.* GONG. A team was formed to prepare a proposal to NSF, with John as leader and project scientist. I became the GONG instrument scientist assisted with the help of many people from both Sac Peak and Kitt Peak. John was careful in planning this project to include many community experts in various advisory and participatory roles, making GONG a truly community project. A proposal was prepared and submitted to NSF in September 1984. Some revisions were requested by NSF and the project came up for review by the NSF Astronomy Advisory Committee of which I happened to be a member at the time. Of course, I was excluded from voting on my own organization's project, but GONG got a green light anyway. By the end of 1985 a technology development proposal had been approved by NSF and funding started to flow. Little did I know that GONG would consume the next decade of my career! The history of the development of the GONG project is recorded in a series of newsletters available at https://gong.nso.edu/info/newsletters. Here I give some personal recollections that stand out to me during my decade of work on GONG.

Once the scientific requirements were transformed into observation instrument parameters, the next step was to decide which of six possible instrumental approaches would be most viable. John astutely arranged a meeting of outside experts in instrumentation at which proponents of the three most viable approaches presented their best arguments. Afterword the committee recommended to use a Michelson interferometer as a Doppler analyzer following previous work by Beckers and Brown (1978). It was then my job to lead the design of an appropriate instrument and to build a prototype. Many people were involved in this effort, but Rob Hubbard, Dick Dunn, and Raleigh Drake were especially valuable members of an excellent design team. A local optical company was selected to build a prototype interferometer. But the skilled key optical person did not devote sufficient attention to the job. The result was a year-long delay and an unfortunate legal battle.

The instrumental technical challenges were significant and time-consuming, and some compromises had to be made. One was to abandon a plan to provide an absolute wavelength reference using a highly stable laser. It turned out to be too hard to feed the reference beam into the instrument with adequate stability. Another difficult problem was dealing with multiple beam paths within the interferometer. Although faint, the effect is to add additional, spurious interfering paths that affect the Doppler measurements. The solution was a redesign of the interferometer and a careful calibration of the instrumental Doppler pattern. Rob Hubbard came up with a key step in this calibration, without which the instrument would not have worked to specification.

By early 1989 the prototype instrument was producing data well enough to convene an in-depth review by outside experts prior to purchasing all the components for the field stations. The review was successful and major construction started. My construction role included assembling and testing ten birefringent filter – interferometer units. I had previous experience with Lyot filters and that was valuable in the production effort.

A few years after the initial data from GONG proved to be successful, some instrumental improvements were proposed. The major ones were to replace the original CCD cameras with 1024 x 1024 pixel CCD cameras for better spatial resolution, replace the original data acquisition system, and to obtain magnetograms continuously. These improvements were tested in 1999 and were deployed by 2001. The Doppler measurements were much improved and the magnetograms have proven to be quite useful. After a decade-long reduction of my time to do science, I was happy to be involved in two research results from GONG. Namely, the unambiguous and frequent observations of magnetic field changes with solar flares (Sudol and Harvey, 2005) and the discovery of ubiquitous, nearly horizontal, rapidly changing magnetic fields in the quiet photosphere (Harvey *et al.*, 2007). There was abundant prior evidence of the latter phenomenon, but for years it was overlooked by me and others.

### 11.3. SOLIS

During one of the periodic budget problems in the mid-1990s, NOAO management thought that the NSF might respond favorably to a proposal for one-time facility investments that would reduce long-term operation costs. The NSO staff was not optimistic about its own chance of success since the NOAO proposal would include two nighttime facility projects in addition to one solar project and NSF might, at best, select only one. Nevertheless, the NSO staff discussed the opportunity and in March 1995, building on long-range strategic studies, proposed to replace several of NSO's aging synoptic observing systems with a suite of four modern instruments. We were astonished when NSF selected SOLIS. Excitement was short-lived as NSF reduced the budget request. This forced the removal of one instrument and cutting corners on the remaining ones.

As in the case of GONG, a large group of dedicated scientists, engineers, and technicians built SOLIS while I served as project scientist. Christoph Keller played major roles in the proposal, design, and initial operation of the SOLIS instruments. His deep intuition about instrumentation was essential. Mechanical engineering was led by Mark Warner, who is presently the project manager for DKIST. Much of the instrument engineering and development was done by David Jaksha and Kevin Schramm. Carole Leiker and Lonnie Cole created most of the real-time software. Carl Henney wrote the initial data reduction programs, which were crucial for the early development of SOLIS science. The SOLIS management structure was very complicated since NOAO's director was the official principal investigator but was not a solar scientist. Additionally, the SOLIS project manager applied for and was accepted as the second DKIST project manager, leaving SOLIS without one for several years, which greatly slowed its development.

The design of the SOLIS vector spectromagnetograph (VSM) was based on a custom CCD that was thought to be well within the state-of-the-art. However, the contracted supplier ultimately failed to produce the device, leading to a long delay in finding a replacement and a reduction in funds available for all the SOLIS instruments. Nevertheless, SOLIS observations started in July 2003 and the VSM replaced the KPVT SPMG after a month of overlap observations.

This is not the place to describe the challenging development of SOLIS and its subsequent operations. At a recent meeting, Sparn *et al.* (2020) listed the elements needed to develop a successful scientific instrument project: partnership, trust, mutual support, leadership, focused goals, defined requirements, science buy-in, and consistent financial support. Upon reflection, SOLIS has not always enjoyed all these elements. Despite this, SOLIS has produced useful and interesting results.

11.4. *SOLAR PHYSICS* JOURNAL

A summary of the first 50 years of this journal may be found at https://link.springer.com/article/10.1007/s11207-016-1022-y. The founding European editors, C. de Jager and Z. Švestka, were joined in 1987 by Robert Howard who became the first of three editors based in the United States (see Figure 12). Oddbjørn Engvold took over from Kees de Jager in 1996. Later that year, Bob Howard asked me to briefly substitute for him during a short absence. He showed me how to deal with incoming submissions, select referees, correspond with the authors, and handle the papers after acceptance. I found the editorial process to be rather antiquated since it dealt with records kept on cards and involved mailing of manuscripts to and from authors, referees, and editors. Bob had automated some of the mundane editorial tasks through a series of VAX Fortran programs that he wrote. I was happy when Bob returned.

As Bob's retirement approached in 1997 he again asked me to fill in for him and to consider if I wanted to become his replacement (subject to approval by the publisher and others). By now I understood the editorial process and it seemed less daunting, and so I agreed, after getting approval from the NSO director. One new task was to proof-read and correct the use of English in all of the accepted papers before they were sent to the typesetters. Since my own language skills are manifestly limited, this was a hard job. Zdeněk, Oddbjørn, and I divided submitted papers according to our respective areas of expertise and through frequent e-mail exchanges. Selecting and interacting with referees was almost always a very satisfying part of the job. I am forever grateful to many colleagues who selflessly agreed to referee papers. Oddbjørn tired of the primitive tools we had available to assist with editing and prevailed upon a colleague to write some modern applications to ease the editing chores. This helped a lot. One of my other tasks was to contact people selected by a small committee to write memoirs such as this one. This was very discouraging to me because in several cases the selected memoirists passed away before completing their writing. I felt like a bad omen and resisted contacting any more people to write memoirs. This morbid sense is probably, in part, responsible for a several-year delay in finishing this memoir.

Toward the end of 2005, the three editors each faced changes in their lives and we decided to resign together in order to give the publisher a chance to freshen the journal with new faces. I asked John Leibacher if he would be interested in being a candidate for editor and he agreed but for no longer than five years. John was selected as an editor and like the other recent editors has done an outstanding job for the community. His five years have since tripled.

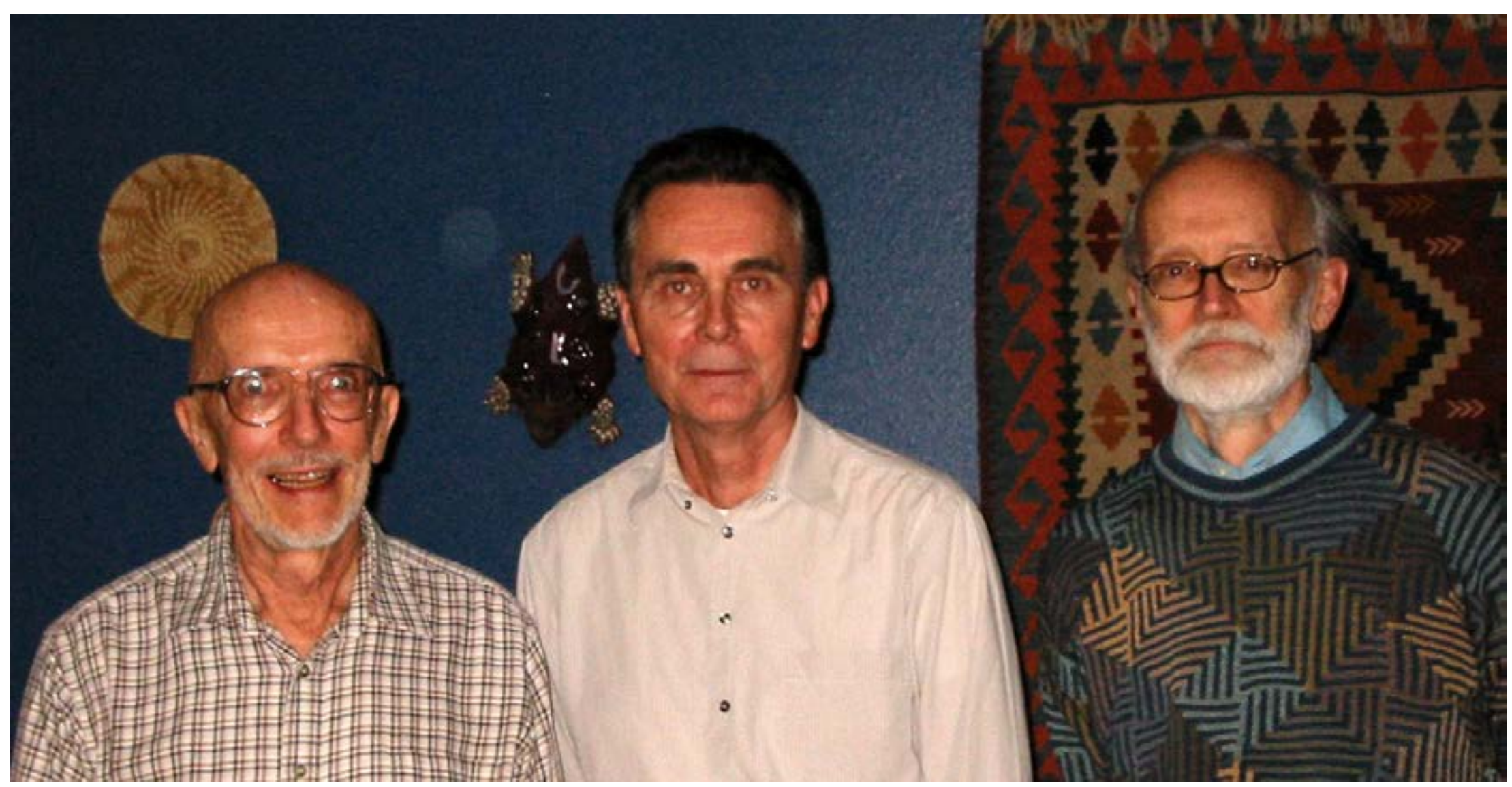

**Figure 12** The progression of US-based editors of *Solar Physics* in May 2005. (Left to right) R. Howard (1987-1997), J. Harvey (1998-2005), J. Leibacher (2005-).

### 12. More life with Karen

Thanks to Karen's former Lockheed employer, Loren Acton, she became a co-investigator on Lockheed's SXT part of the Yohkoh mission. This led to several trips to Japan, sometimes with me as accompanying spouse. Helioseismology meetings in Japan reversed our roles, and as a result we met a wide range of wonderful friends and colleagues in Japan. One such event is shown in Figure 13.

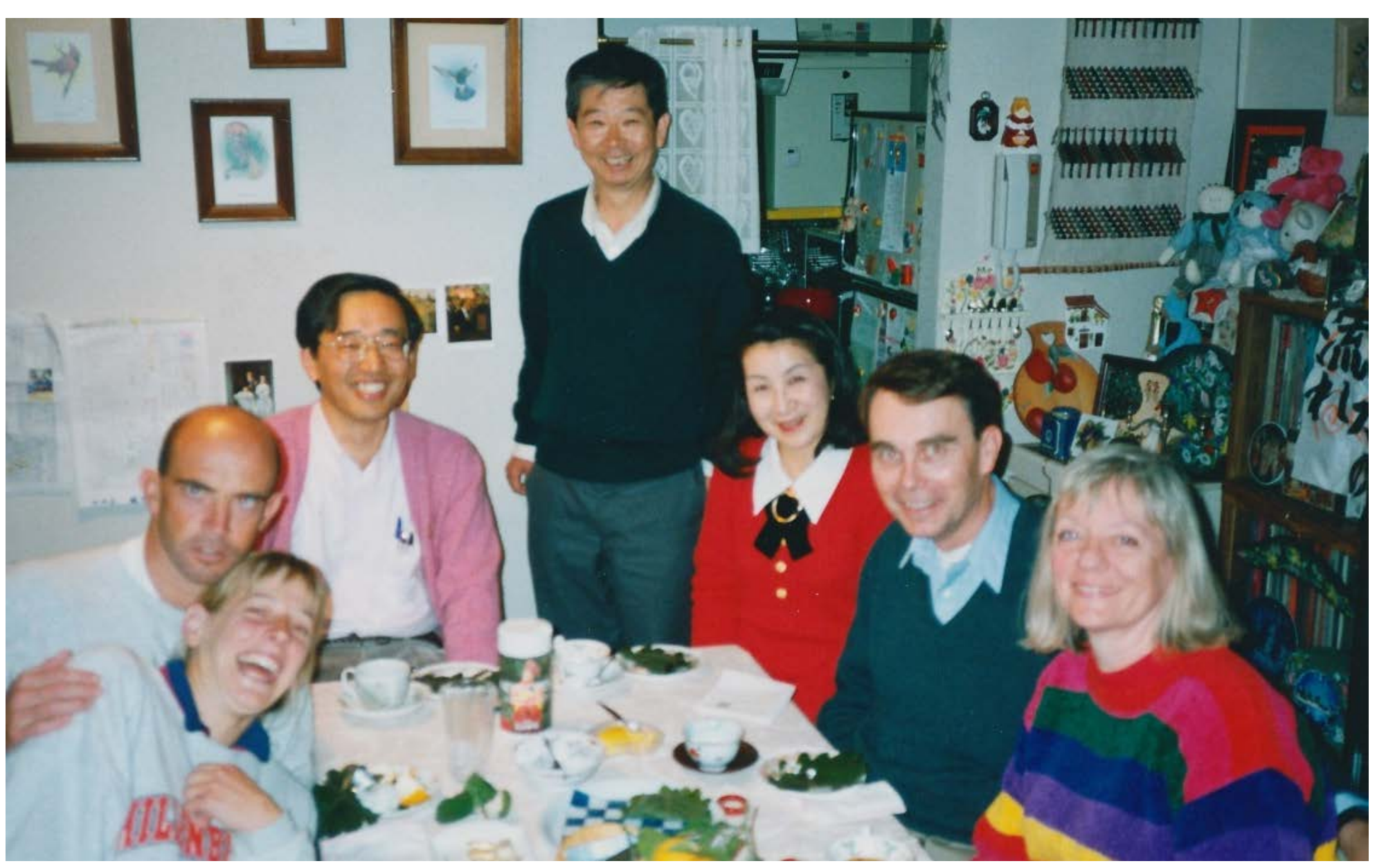

**Figure 13** Dinner at the Shibahashi home. (Left to right) Stuart and Colleen Jefferies, H. Shibahashi, Y. Osaki, Mrs. Shibahashi, me, and Karen. (Photo by Shiro Shibahashi.)

Thanks to Kees Zwaan of Utrecht University, Karen started a PhD thesis in 1984, thus fulfilling a long-desired wish to earn a doctoral degree. It was a massive study of the magnetic properties of active regions and ephemeral regions and was a highlight of her career (Harvey-Angle, 1993). Karen's thesis defense took place in Utrecht with Karel Schrijver and me acting as her supporting paranymphs in front of a distinguished panel of splendidly-robed judges chaired by Kees de Jager. Karen's family was present, the traditional ceremony went well, and a fine time was had by all.

Our lives and careers continued happily intertwined until everything changed when Karen was diagnosed with cancer in 1998. Her treatments were frequent and debilitating but Karen continued cheerfully working as a welcome distraction until her last week in 2002. Karen's obituaries provide details about her career (*e.g.* https://aasjournals.github.io/aas-obits-mirror/karen-lorraine-harvey-1942-2002.html). I fell into serious depression and my work suffered. John Leibacher suggested to me that an appropriate tribute to Karen would be to finally initiate a long-desired early career award of the Solar Physics Division of the American Astronomical Society. There were already some funds donated for this purpose, and I contributed enough to sustain the Karen Harvey early career award for at least 25 years. It has been wonderfully satisfying to see a selection of outstanding young researchers receive the award and become leaders in our field (https://spd.aas.org/prizes/harvey/previous).

**13. National Solar Observatory under AURA 2010-2020**

NSO was a small part of NOAO and suffered from low priority and poor success in competition for resources. After many years NSO finally became independent in 2010 as I turned age 70. As a result of NSF senior reviews (with little input from the solar community), the tasks of NSO became to divest existing NSO facilities on Kitt Peak and Sacramento Peak, complete the DKIST project, move NSO to new offices in Boulder, Colorado, and keep the synoptic observing program alive. These tasks have largely been accomplished through the leadership of NSO director Valentin Martinez-Pillet. However, the latter task has been challenging due to an NSF cap on funding. By demonstrating the value of GONG and SOLIS magnetograms for space weather purposes, some extra funding for the synoptic program was obtained by Frank Hill. However, NSO science research based on the synoptic observations now depends on competitive funding opportunities and the synoptic program is greatly stressed. Plans for a new generation of synoptic instrumentation are being developed throughout the international solar community.

## 14. Life with Terrie

My long-time mentors and friends Bill and Dorothy Livingston were concerned about my serious depression after Karen's passing. Luck favored me once again when they introduced me to Terrie Fernandez, and she put up with my awkward return to a social life. Terrie introduced me to the fascinating microscopic realm through her career work in cytology and brought art into my life through her artistic talents and interests. My life brightened, though I have a permanent loss after Karen's passing. After two years of companionship, we married in 2006. Terrie has traveled with me to many solar meetings and likes to socialize with members of our community and their spouses. One such example is Figure 14. She also introduced me to many of her friends, which greatly broadened my social interactions.

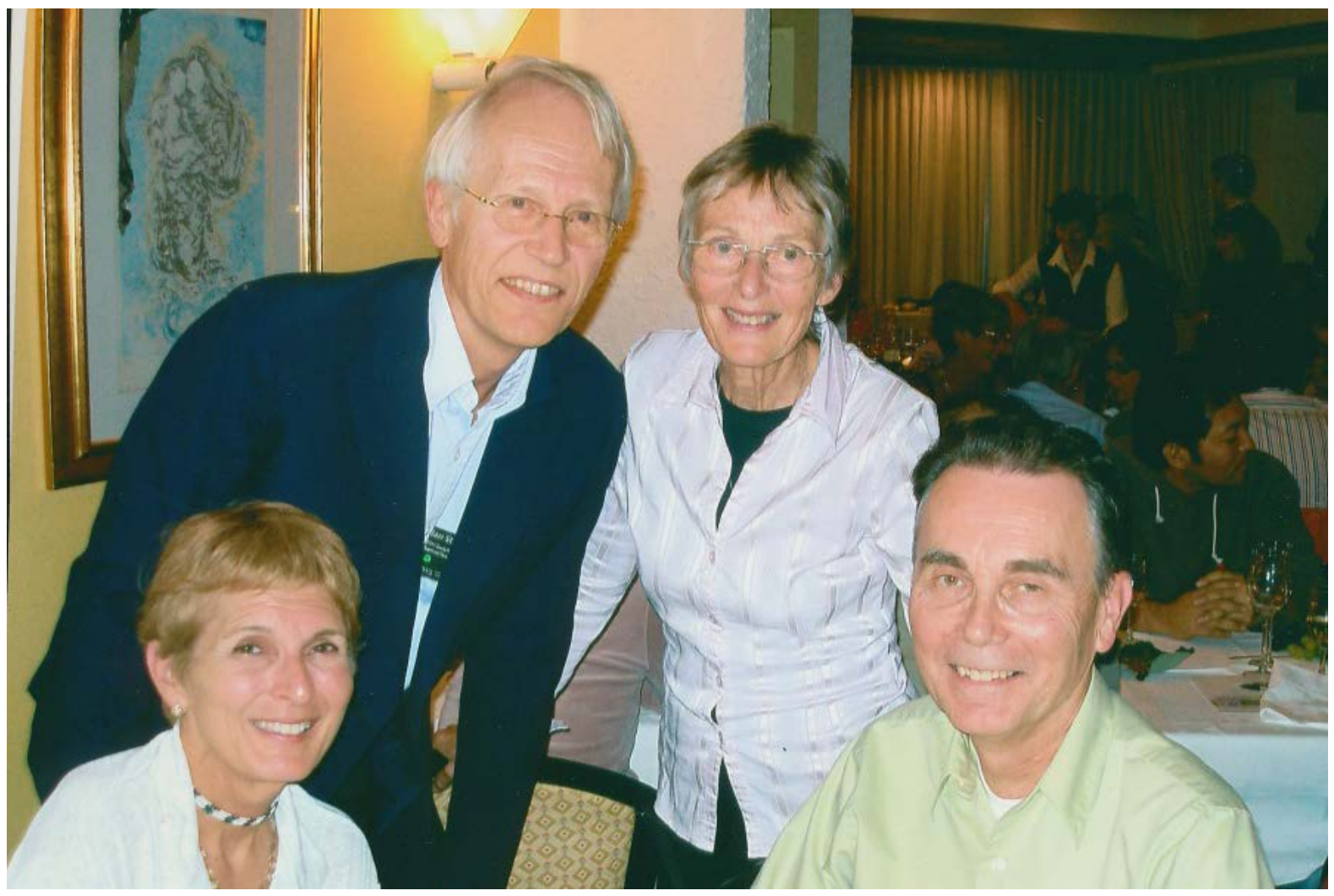

**Figure 14** (Left to right) My wife,Terrie Fernandez, long-time friends Jan and Joyce Stenflo, and me at the celebration of Jan's career at Ascona in September, 2007.

## 15. Finale

Writing this memoir made me face the reality that my professional career is nearly completed. I fully realize and greatly appreciate that whatever successes I've had stem from the great good luck and generous people and mentors that have touched my life. It is satisfying to have largely realized my goal of working at NSO in a privileged job to serve the community by both direct support of and development of facilities for the community to study the Sun. When my name appeared on an Astronomy Acknowledgements Index (Verner, 1991) it indicated gratifying progress in reaching these goals.

It has been rewarding to help build and sustain productive facilities including the Kitt Peak Vacuum Telescope and its 512-channel magnetograph and spectromagnetograph; spectropolarimeters for the McMath-Pierce Telescope main spectrograph and Fourier transform spectrometer; early solid-state detector arrays for helioseismic spectroscopy; imaging instruments for helioseismology at South Pole; the instrument for the Global Oscillation Network Group; and the SOLIS instruments. I emphasize again that all of these projects depended on the skill and dedication of many other people.

My science contributions have almost always been collaborative and intended to demonstrate and explore capabilities of new equipment rather than targeted efforts to solve key problems. I'm happy to have participated with much help and collaboration, in making the first Zeeman effect measurements of the coronal magnetic field and images of the chromospheric magnetic field (Harvey 1963, 1969; Harvey and Hall, 1971), in early potential field extrapolations of coronal magnetic fields (Newkirk, Altschuler, and Harvey, 1968), in the gradual discovery of the internetwork magnetic field (Livingston and Harvey, 1971b; 1975), the first inversions of spectropolarimetric line profiles (Harvey, Livingston, and Slaughter, 1972), instrument setups that helped unveil the second solar spectrum (Stenflo, Twerenbold, and Harvey, 1983), filling the intermediate degree gap in the solar oscillation spectrum (Duvall and Harvey, 1983), first helioseismic inversions of rotation throughout most of the solar interior (Duvall *et al.*, 1984), discovery of the rapidly changing ubiquitous horizontal magnetic field (Harvey *et al.*, 2007), and providing unequivocal evidence of magnetic field changes associated with a large number of flares (Sudol and Harvey, 2005).

My major career failure was starting too many projects and not finishing them. I also missed several big opportunities. One example was a discrepancy between network longitudinal magnetic fields measured with a Babcock-type magnetograph using spectrum lines with different Zeeman splitting. I noticed a factor of 2.5 discrepancy during my 1966 thesis observations at KPNO but thought that I had used a 2.5 gain switch setting on an amplifier and incorrectly recorded it. Stenflo (1973) got it right, and I was happy to help him use the KPNO equipment in his further work in this area. Another miss was a 1972 project with John Leibacher and O. R. White to measure the spatial-temporal spectrum of solar five-minute oscillations. The equipment and weather misbehaved but the main fault was my poor understanding of the best way to do the 2D spatial scans. Franz Deubner (1976) knew better. An inexcusable miss was overlooking for many years indications of a ubiquitous, rapidly changing, and mostly horizontal magnetic field in the photosphere. This finally became so obvious that it could not be ignored. Our discovery paper (Harvey *et al*., 2007) was almost immediately followed and consequently superseded by higher resolution findings from Hinode (Lites *et al*., 2008).

I've recently enjoyed studying the history of solar research and hope to publish some reminders of its lesser known aspects. John Briggs and I are doing an in-depth study of Lewis Rutherfurd's claim to have photographed solar granulation six years before Janssen's well known results. This study, thanks to the discovery of his original 1871 plates, confirmed Rutherfurd's claim and led us to other little-known precursors of Janssen's work (Harvey, Briggs, and Prosser, 2017). It was also fun to learn about Arago's pioneering 1811 solar polarimetry and to build a replica of the first solar polarimeter (Harvey, 2015). This involved reading fascinating French journals dating back to revolutionary times. Hooray for the Internet!

In my mostly retired status, my main goal is to restore, curate, and document the archive of thousands of full-disk magnetograms obtained at Kitt Peak from 1970 through 2003. Andrés Munoz-Jaramillo is trying to keep me focused on this goal. But my 50+ year fascination with chromospheric magnetic fields continues with a recent addition to the SOLIS Vector SpectroMagnetograph (VSM) of full Stokes spectropolarimetry of the chromosphere (Harvey *et al*., 2016). The team working on the initial data has produced results that are an exciting distraction from the archival work. It was fascinating to discover that scattering linear polarization is ubiquitous across the solar disk in the core of the 8542 Å line (Harvey and SOLIS Team, 2020).

In whatever spare time comes my way I will try to finish some of my many unfinished projects. In a return to my amateur origins, I've tried to convince some of today's highly advanced amateurs that one of them can be the first to do successful amateur helioseismology. That would be a most satisfying finale to my very lucky career of deconstructing sunlight.

**Acknowledgements** I'm truly indebted to the hundreds of people I've been privileged to know and work with, and I wish continued success to the international community of solar researchers. I thank my wife, Terrie, and our friendly next-door neighbors, English professors Tilly and John Warnock, and an (almost) anonymous referee for many suggestions that greatly improved this memoir.